\documentstyle[aps,prc,preprint,epsf]{revtex}
\begin{document}

\title{
Real Time Dynamics of Colliding Gauge Fields and the ''Glue Burst''  
}
\vspace{0.1cm}
\author{ W. P\"oschl and B. M\"uller \\
Department of Physics, Duke University, Durham, 
NC 27708-0305, USA }
\vspace{0.2cm}
\maketitle
%
%

\begin{abstract}
The Yang Mills equations provide a classical mean field description
of gauge fields. In view of developing a coherent description of the
formation of the quark gluon plasma in high energetic nucleus-nucleus 
collisions we study pure gauge field dynamics in 3+1 dimensions.
In collisions of wave packets, numerically simulated on a SU(2) gauge
lattice, we study transverse and longitudinal energy currents.
For wave packets with different polarizations in color space,
we observe a time delayed fragmentation after the collision resulting 
in a rapid expansion into transverse directions. 
We call this phenomenon the ''glue burst''. An analysis of the 
Yang Mills equations reveals the explanation for this behavior.
We point out that this effect could play a role in
ultra-relativistic heavy-ion collisions.
\end{abstract}
\vspace{1.0cm}
%
%
%
%
\section {Introduction}
%
%
\noindent
It is one of the most challenging topics in the theory of
ultra-relativistic heavy-ion collisions to develop a coherent
description of the formation of the quark gluon plasma.
Many descriptions of the time evolution of such collisions have
been developed in the past years 
\cite{Bass.98,Faessler.98,Venu.97,Wang.96,Geiger.95}. 
Some of the most
recent are the ultra-relativistic quantum molecular dynamics
model (UrQMD) \cite{Bass.98} and the parton cascade model \cite{Geiger.92}.
While the UrQMD has been developed for collisions 
at center of mass energies below $10\,{\rm GeV/u}$, 
the parton cascade model is valid for energies at and above 
$100\,{\rm GeV/u}$ where the transition into the plasma
phase is expected to occur. 
Both are based on the idea of a perturbative scattering
of particles within transport models and describe in
principle the evolution of a collision from the first
contact of the nuclei throughout the high density phase 
until to the point where the last Hadrons are freezing out. 
These models however contain still a variety of problems.
One of the problems concerns the description of the 
initial state of the colliding nuclei.
The transport equations start from probability distributions of
partons in the phase space. In reality however, the states of
the nuclei are described through coherent parton wave functions.
The incoherent parton description especially breaks down 
at exchanges of small transverse momenta.
A few years ago, 
McLerran and Venugopalan proposed \cite{McLerran.94} that the proper
solution of these difficulties is the perturbative expansion not around
the empty QCD vacuum but around a vacuum of the mean color fields
which accompany the quarks in the colliding nuclei. This idea motivates
the development of a combination of the parton cascade model 
\cite{Geiger.92} with a coherent description of the 
initial states. This difficult goal can be accomplished in
small steps through systematic studies. Recently, we have
proposed a combination of a gauge lattice description for
the soft color fields with a transport model for color
charged particles \cite{BMP.98}. Leaving out the collision terms first,
this model then has been applied to simulate the collision
of clouds of color charged particles accompanied by soft
color fields in 3+1 dimensions \cite{PM.98}. The field
energy distributions obtained for times shortly after the
collision have shown transverse energy flows
resulting from glue field scattering in the center of collision.
For times larger than $1\,{\rm fm/c}$ glue field radiation
seemed to be dominant. The sudden appearance of transverse 
energy flows during the overlap time of the nuclei was the
motivation to study here the pure glue mean field dynamics leaving
out the particles. The time evolution of colliding 
Yang Mills field wave packets has been studied a few years ago in 1+1 
dimensions \cite{Hu.95}. These calculations have shown
that wave packets of equal polarization in color space
(Abelian case) 
do not interact whereas wave packets polarized into different
color directions (non-Abelian case) 
decay after the collision into low frequency
modes of high particle multiplicity in the sense of 
Ref. \cite{Ring.91,Esp.91}. This mechanism of a coupling
between high frequency short wavelength modes and low
frequency long wavelength modes in the Yang Mills equations
has also been observed in Ref. \cite{Gong.94}. 

In the present paper, we focus on the transverse dynamics and the
coupling between longitudinal and transverse energy flows
in collisions of Yang Mills fields. A study of the transverse
dynamics requires simulations in at least 2+1 dimensions.
Subsequently, we describe the method used to solve the
Yang Mills equations and present results obtained from collisions
simulated in 3+1 dimensions on a SU(2) gauge lattice. These
studies also reveal an interesting behavior of the time-evolution
of non-Abelian gauge fields. 
%
%
%
%
\section {Time Evolution on the Gauge Lattice}
%
%
In the Lie algebra LSU(2), we define the adjoint gauge
fields ${\cal A}^{\mu}(x) := A_c^{\mu}(x)T^c$ 
and the adjoint
field strength tensor ${\cal F}^{\mu\nu} := F^{\mu\nu}_cT^c$.
Einsteins sum convention has to be applied in the Euclidean metric  
for upper and lower color indices and in the Mincowski metric
for upper and lower greek indices.
The symbols $T^c$ with color index $c=1,2,3$ denote the
generators of LSU(2) obeying the commutation relations
$\bigl[T^a,T^b\bigr]_- = if_{abc}T^c$ and hence 
${\cal A}^{\mu}(x), {\cal F}^{\mu\nu}\in$ LSU(2)
for all $x\epsilon{\bf R}^4$. Here, we chose the representation
$T^c= 1/2\,\sigma^c$ with the Pauli matrices $\sigma^c$.  
Further below, we use also $T^0:={1\over 2}{\bf 1}_2$ which is
linearly independent from the generators $T^c$.
With these conventions,  
we denote the Yang Mills equations in the short form 
\begin{equation}
\label{Eq.1}
\bigl[{\cal D}_{\mu},{\cal F}^{\mu\nu}\bigr]_- = 0,
\end{equation} 
where ${\cal D}_{\mu}$ is defined 
\begin{equation}
\label{Eq.2}
{\cal D}_{\mu} := \partial^{\mu} - ig{\cal A}^{\mu}. 
\end{equation}
With this definition of the covariant derivative 
${\cal D}_{\mu}$ on the SU(2) mainfold, 
and with the definitions 
${\cal E}^{\mu}(x) 
:= E_c^{\mu}(x)T^c$, 
${\cal B}^{\mu}(x) := B_c^{\mu}(x)T^c$ 
(${\cal E}^{\mu}(x),{\cal B}^{\mu}(x)\,\in$ LSU(2))  
of the adjoint color electric and color magnetic field
quantities, the Yang Mills equations Eq. (\ref{Eq.1})
can be expressed in a form which is similar to the 
U(1) Maxwell equations in the vacuum  
\begin{eqnarray}
\label{Eq.3}
\bigl[{\vec {\cal D}},{\vec {\cal E}}\bigr]_- &=& 0, \\ 
\label{Eq.4}
\bigl[{\vec {\cal D}},{\vec {\cal B}}\bigr]_- &=& 0, \\ 
\label{Eq.5}
\bigl[{\vec {\cal D}}\times, {\vec {\cal E}}\bigr]_- &=& 
\bigl[{\cal D}_0,{\vec {\cal B}}\bigr]_- , \\ 
\label{Eq.6}
\bigl[{\vec {\cal D}}\times,{\vec {\cal B}}\bigr]_- &=& 
\bigl[{\cal D}_0,{\vec {\cal E}}\bigr]. 
\end{eqnarray}
With the condition $\bigl[ {\cal A}(t,\vec x),{\cal A}(t,\vec x')\bigr]_- = 0$
for all $\vec x,\vec x'\in {\bf R}^3$ for one arbitrary real time $t$, 
the equations (\ref{Eq.3}) to (\ref{Eq.6}) are indeed identical with 
the Maxwell equations. In this so called Abelian (or U(1)) case, 
one expects a linear behavior of the solution and it therefore 
provides an important test through comparison with the solution 
in the general non-Abelian case.
The Yang Mills equations can be solved in an efficient 
manner on a gauge lattice in a Hamiltonian framework where
we choose the temporal gauge ${\cal A}^0 = 0$.

A lattice version of the continuum Yang Mills equations
is constructed by expressing the color field amplitudes as 
elements of the corresponding Lie algebra, i.e. ${\cal E}_{{\vec x},k},
{\cal B}_{{\vec x},k}\,\in$ LSU(2) at each lattice site $\vec x$.
We define the following variables.
\begin{eqnarray}
\label{Eq.7}
{\cal U}_{{\vec x},l}
&=& \exp(-iga_l{\cal A}_l(x)) \,\,=\,\,
{\cal U}^{\dagger}_{{\vec x}+l,-l}     \\
\label{Eq.8}
{\cal U}_{{\vec x},kl}
&=& {\cal U}_{{\vec x},k}\, {\cal U}_{{\vec x}+k,l}\,
    {\cal U}_{{\vec x}+k+l,-k}\, {\cal U}_{{\vec x}+l,-l}
\end{eqnarray}
In adjoint representation the color electric and color magnetic
fields are expressed in terms of the above defined
link variables ${\cal U}_{{\vec x},l}$ and plaquette variables
${\cal U}_{{\vec x},kl}$ in the following way.
\begin{eqnarray}
\label{Eq.9}
{\cal E}_{{\vec x},j} &=& { {1}\over{iga_j}}\,\dot {\cal U}_{{\vec x},j}
                   {\cal U}^{\dagger}_{{\vec x},j}    \\
\label{Eq.10}
{\cal B}_{{\vec x},j} &=& { {i}\over{4ga_ka_l}}\,\epsilon_{jkl}\,
\bigl({\cal U}_{{\vec x},kl} -
{\cal U}^{\dagger}_{{\vec x},kl} \bigr).
\end{eqnarray}
The lattice constant in the spatial direction $l$ is denoted by $a_l$.
As one can see from (\ref{Eq.6}), the gauge field ${\cal A}_{{\vec x},l}$
is expressed in terms of the link variables
${\cal U}_{{\vec x},l}\,\epsilon $ SU(2), 
which represent the parallel transport
of a field amplitude from a site $x\,\epsilon\, X$
to a neighboring site $(x+l)\,\epsilon \, X$ in the direction $l$.
We choose ${\cal U}_{{\vec x},i}$ and ${\cal E}_{{\vec x},i}$ as the
basic dynamic field variables and
numerically solve the following equations of motion.
\begin{eqnarray}
\label{Eq.11}
\dot{\cal U}_{{\vec x},k}(t) &=& 
i\,g\,a_k\,{\cal E}_{{\vec x},k}(t)\,{\cal U}_{{\vec x},k}(t)
\\
\label{Eq.12}
\dot{\cal E}_{{\vec x},k}(t) &=& {i\over{2ga_1a_2a_3}} \sum\limits^3_{l=1}
\Bigl\{{\cal U}_{{\vec x},kl}(t) -
{\cal U}^{\dagger}_{{\vec x},kl}(t)
\,\Bigr.
\nonumber \\ 
&-&  \Bigl. {\cal U}^{\dagger}_{{\vec x}-l,l}(t)\,
\Bigl( {\cal U}_{{\vec x}-l,kl}(t) -
{\cal U}^{\dagger}_{{\vec x}-l,kl}(t)\Bigr)\,
{\cal U}_{{\vec x}-l,l}(t)\, \Bigr\}.
\end{eqnarray}
%
%
%
%
\section {Numerical Results}
%
%
\noindent
First, we study the collision of plane wave packets which have
initially constant amplitude in the transverse planes of a
3-torus lattice implying periodic boundary conditions.
The size is chosen 20x20x2000 lattice points. We denote
the lattice spacing in the space direction $l$ by $a_l$.
For a collision, we initially arrange wave packets of a
Gaussian shape in the longitudinal direction.
The average momenta of the wave packets are 
$(0,0,\pm\overline k_3)$ and their momentum spread is denoted
$\Delta k_3$. As a consequence of the lattice discretization
these quantities are restricted to values between the maximal 
and minimal Fourier-momenta on the lattice:
\begin{eqnarray}
\label{Eq.18}
k_l^{min} = {\pi\over{(N_la_l)}},\qquad 
k_l^{max} = {\pi\over{(N_la_l)}},
\\
\label{Eq.19}
k_l^{min}\,\ll\, \Delta k_l\, \ll\, \overline k_l\,\ll\, k_l^{max} 
\end{eqnarray}
The polarization in color space is defined by the unit
vectors $\vec n_{\rm L}^c$ for the left (L) and 
$\vec n_{\rm R}^c$ for the 
and right (R) moving wave packet respectively.

\noindent
We express the initial conditions through the gauge fields
\begin{eqnarray}
\label{Eq.20}
A_{L}^c=\delta_{l1}{\vec n}^c_{L}\phi(+t,x_3-Z_{L}), \\
\label{Eq.21}
A_{R}^c=\delta_{l1}{\vec n}^c_{R}\phi(-t,x_3-Z_{R})
\end{eqnarray}
where the scalar function $\phi(t,x_3)$ 
defines the initial wave packet
\begin{eqnarray}
\label{Eq.22}
\phi(t,x_3)&=&\phi_0
\exp{\bigl(-{1\over 2}\Delta k_3^2(t+x_3)^2\bigr)}
\cos{\bigl(\overline k_3(t+x_3)\bigr)}, \\
\label{Eq.23}
\phi_0 &=&
\sqrt{ {(2\Delta k_3)}/({\sqrt{\pi}\sigma\overline k_3})}
\bigl( 1+e^{-(\overline k_3/\Delta k_3)^2 } \bigr) 
\end{eqnarray}
The amplitude factor $\phi_0$ is determined for a 
normalized wave packet when $\sigma = 100a_1a_2$.
The parameter $\sigma$ is used to control the amplitude
and allows to adjust the energy contained in a 
wave packet. In a particle interpretation it describes
the total cross section per particle contained in the
wave packet \cite{Hu.95}. The second term in $\phi_0$
is usually negligibly small for wave packets since 
$\overline k_3/\Delta k_3 \gg 1$.
We chose the polarizations in color space
$\vec n_R=\vec n_L = (0,0,1)$ for the
Abelian case and $\vec n_R = (0,0,1),\,\vec n_L = (0,1,0)$
for the non-Abelian case.
Once the initial fields are
mapped on the lattice, the time evolution of the collision
starts from the superposed initial conditions
\begin{eqnarray}
\label{Eq.24a}
{\cal U}^{(0)}_{{\vec x},l} &=&
{\cal U}^{(0)}_{{\vec x},l,L}\cdot{\cal U}^{(0)}_{{\vec x},l,R},
\\
\label{Eq.24b}
{\cal E}^{(0)}_{{\vec x},l} &=&
{\cal E}^{(0)}_{{\vec x},l,R}+{\cal E}^{(0)}_{{\vec x},l,L},
\end{eqnarray}
respectively.
A linear superposition of solutions obeys the Yang Mills
equations only if these solutions have no overlap . Therefore,
the initial separation $\Delta Z$ 
should be much larger than $1/{\Delta k_3}$.

\noindent
The calculation contains four parameters.
The relative color polarization which is parameterized by
the angle $\theta_{\rm C}$ defined through 
$\vec n_L\cdot \vec n_R = \cos(\theta_{\rm C})$. 
The average momentum of the wave packets
$\overline k$ and their width $\Delta k_3$ and the coupling
constant $g$ which can be rewritten in terms of the parameter
$\sigma$ as $g'=g/\sqrt{\sigma}$ by
simultaneously rescaling the field ${\cal A}$ as
${\cal A}'=\sqrt{\sigma}{\cal A}$.
Consequently, the system shows the same dynamics for different
values of $g$ and $\sigma$, as long as the ratio
$g'=g/\sqrt{\sigma}$ is kept fixed.
Multiplying the Eq. (\ref{Eq.41}) in section 4 by a factor $g$, 
shows that each amplitude ${\cal A}^{\mu}$ absorbs a $g$ such 
that no $g$ is left over in Eq. (\ref{Eq.41}). 
For $\overline k_3 \gg \Delta k_3$, the energy contained
in one wave packet is essentially $\overline k_3/\sigma$, i.e.
for given $\overline k_3$ the energy is determined through
the amplitude and vice versa the amplitude is thus fixed by 
the energy. Once the amplitude is fixed, it makes sense
to use the coupling constant as an additional parameter
which determines the reaction dynamics of colliding wave packets.
This is preferable also in view of studies in the future where
normalized Dirac fields or color charged classical particles
shall be included. In such cases $g$ appears also in front of the 
source current of the inhomogeneous Yang Mills equations. 
We therefore keep the explicit denotation of $g$ in the
subsequent sections.

\noindent
In the following we present results from a simulated collision in the
non-Abelian case with the parameters $\overline k_3=\pi/(2a_3)$,
$\Delta k_3=\pi/(100 a)$, $g=1$, $\sigma = 100 a^2 = 1.0 {\rm fm}^2$.
The simulation was performed on a uniform lattice with constants 
$a_1=a_2=a_3=a$ using a time step size $\Delta t= a/10$.
If we would consider a collision of nuclei at low energies,
$a=0.1\,{\rm fm}$ would be an appropriate choice for the
above parameter settings.  
In particluar, $\Delta k_3=\pi/(100 a)$
corresponds thus to a FWHM of $10\,{\rm fm}$ which is
approximately the diameter of a $^{208}Pb$-nucleus.
It is important to note that
for the above chosen values, the calculation runs in
the regime of weak coupling.
Subsequently, we refer to the direction of the collision
axis as the ''longitudinal direction'' and to directions
perpendicular to the collision axis as the ''transverse directions''.
Accordingly, we define the transverse and longitudinal energy densities
of the color electric field
\begin{eqnarray}
\label{Eq.25a}
w_T^{(E)}(t,x_3) &=&\int dx_1dx_2 \sum\limits_{l=1}^2
       {\rm Tr}\bigl({\cal E}_l{(t,{\vec x})}{\cal E}_l{(t,{\vec x})}\bigr),
\\
\label{Eq.25b}
w_L^{(E)}(t,x_3) &=&\int\int dx_1 dx_2
        {\rm Tr}\bigl({\cal E}_3{(t,{\vec x})}{\cal E}_3{(t,{\vec x})}\bigr)
\end{eqnarray}
Fig. 1 displays $w_T^{(E)}$ plotted over the $x_3$-coordinate
at various time steps $t_n$ as indicated on top of the curves.
%
%
%
%
%
%
\begin{figure}[H]
\vspace{1.0cm}
\centerline{
\epsfysize=18cm \epsfxsize=14cm
\epsffile{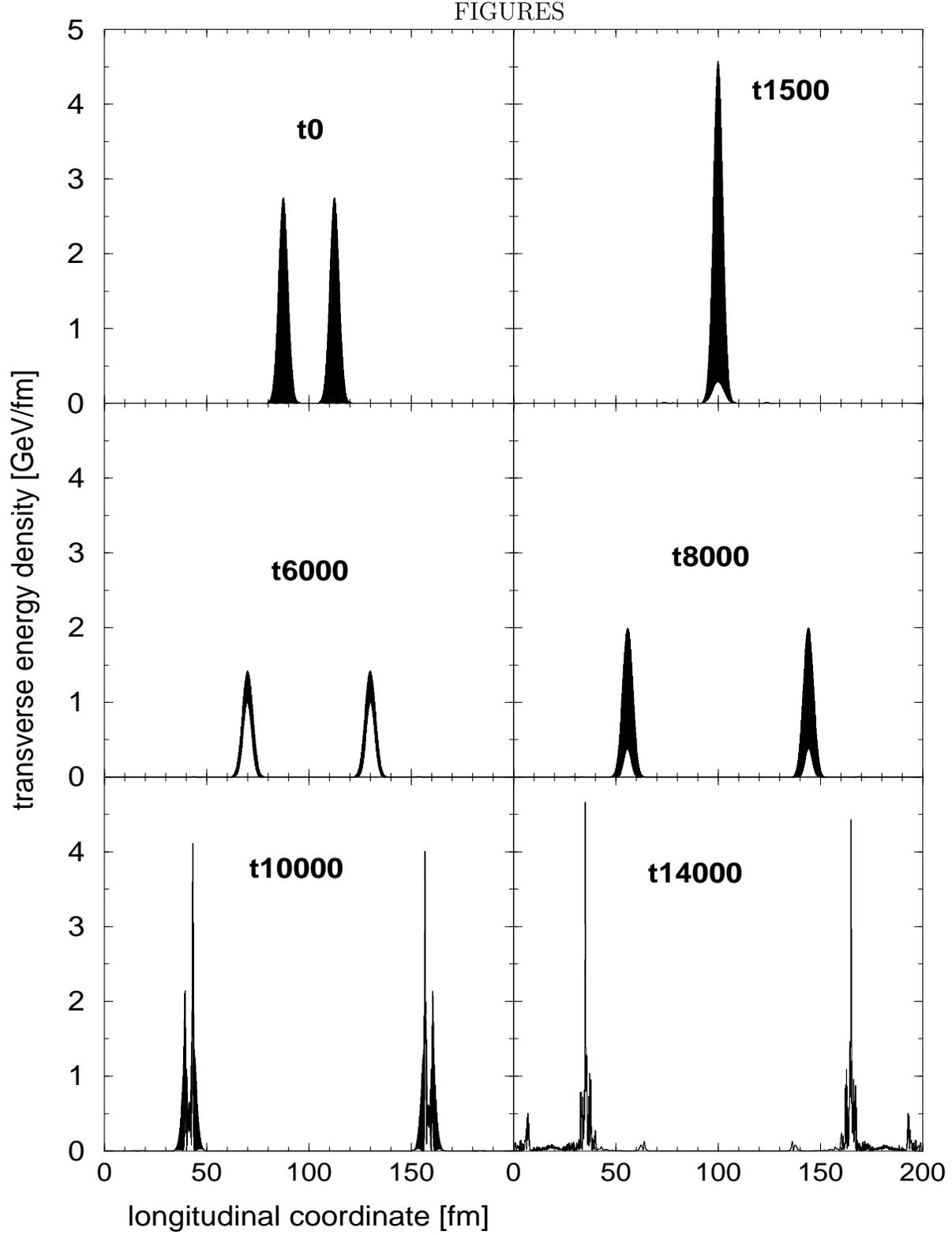}
}
\vspace{1.0cm}
\caption{Transverse energy densities $w_T^{(E)}(t,x_3)$ are
         plotted over the collision axis (longitudinal coordinate) 
         for selected time steps $t_n$.  }
\end{figure}
\noindent
At initial time $t_0$ the distributions $w_T^{(E)}(t_0,x_3)$ 
of the wave packets are completely filled resulting from the strong
oscillations in the longitudinal direction according to 
$\overline k_3=\pi/(2a_3)$.
After 1500 time steps the wave packets are colliding and have reached
maximum overlap. The smooth white zone at the bottom results
from a phase shift between the two superposed waves, i.e.
the maxima of the wave packet (1) to not coincide with
the maxima of the wave packet (2) at time step $t_{1500}$.
After the wave packets have passed through each other 
(at about $t_{2500}$), a small white zone remains in 
the distribution $w_T^{(E)}(x_3)$ of each and keeps
continuously growing. 
At time step $t_{6000}$ ($t=600a$) the large fraction of the initial 
high frequency oscillations is reduced to a small remaining
contribution visible on the surface of the distibutions
in Fig. 1. The hight of the two receding humps decreases
accordingly while the energy carried by each wave packet
is constant in time. 
Almost all the
energy which has originally been carried by short wavelength modes
around $\overline k_3$ has been transmitted into long wavelength 
modes which have filled up the vallies in the oscillating 
distribution $w_T^{(E)}(x_3)$.
This behavior agrees qualitatively with results obtained
in Ref. \cite{Hu.95} where collisions of wave packets
have been studied on a one dimensional gauge lattice
and for times not larger $t=600a$. 
At time step $t_{8000}$ however, 
we observe that energy is partly transmitted back 
into high frequent modes before the wave packets start to decay
around time step $t_{10000}$. At time step
$t_{14000}$, the energy distribution has expanded into
longitudinal direction. The appearance of a circular
polarization of the receding wave packtes could in
princple lead to the same behavior before they decay.
This possibility, however, is excluded by our numerical
results which show - in the case of plane waves -
no excitation of modes with
$x_2$-polarization for all times.
This can be veryfied down to floating point precision
but it has also a fully analytic explanation 
through Eq. (\ref{Eq.41}) in section 4 for $\nu = 2$.

\noindent
This behavior does not appear in a collision of wave packets
which are equally polarized in color space. In this case
the shape of the two humps in the distribution 
$w_T^{(E)}(x_3)$  at the time step $t_{14000}$ 
is almost identical with the shape
at time $t_0$. Deviations result from lattice dispersion.
%
%
%
%
%
%
\begin{figure}[H]
\vspace{1.0cm}
\centerline{
\epsfysize=18cm \epsfxsize=14cm
\epsffile{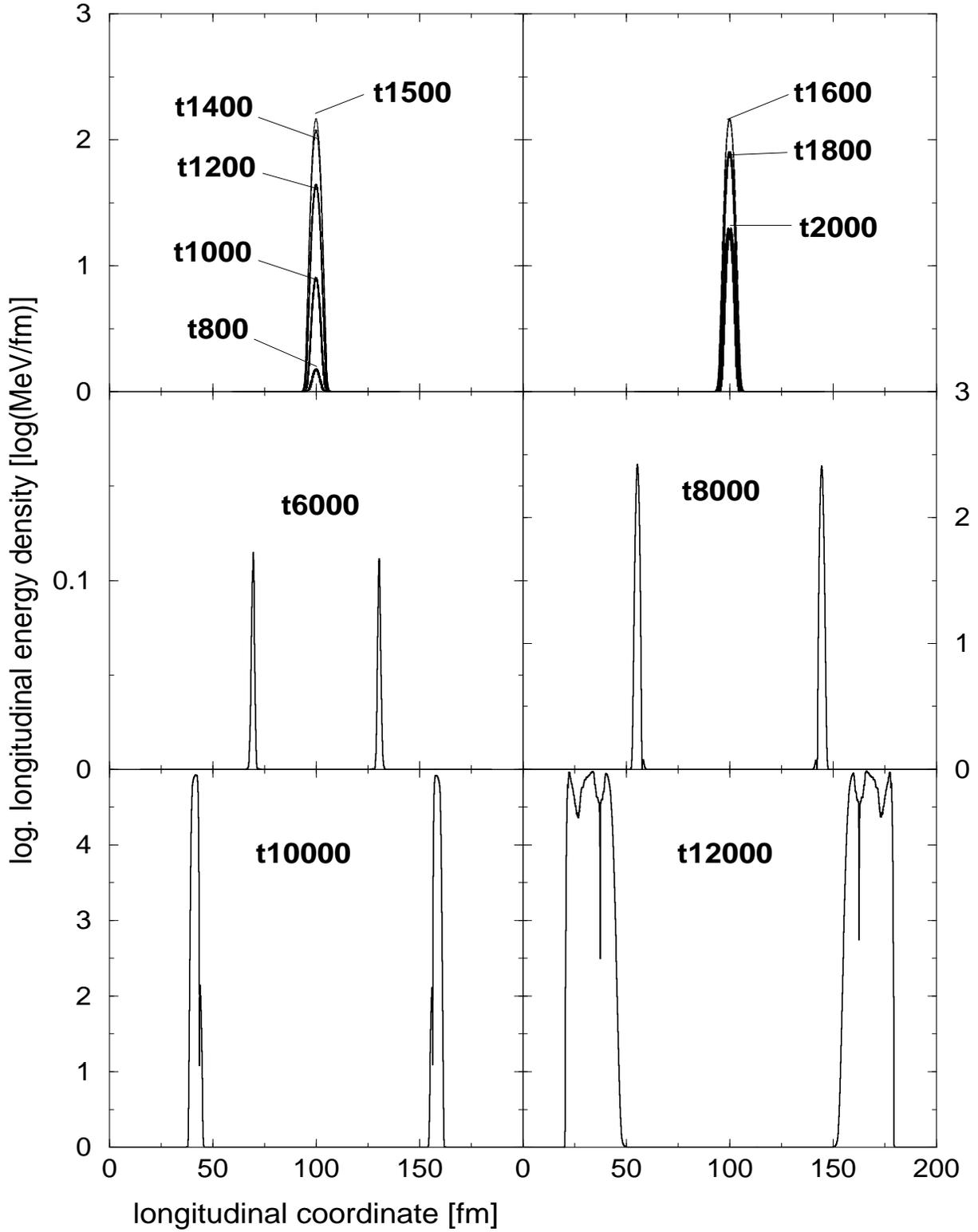}
}
\vspace{1.0cm}
\caption{Longitudinal energy densities $w_L^{(E)}(t,x_3)$ are
         plotted over the collision axis (longitudinal coordinate) for
         selected time steps $t_n$ }
\end{figure}
\noindent
In Fig. 2, the corresponding longitudinal energy densities 
$w_L^{(E)}(t,x_3)$ are displayed on a logarithmic scale 
for selected time steps $t_n$.  
\noindent
We remember that the wave packets were
initially polarized into the transverse $x_1$-direction and
consequently $w_L^{(E)}$ has to be zero as long as
they propagate free. However, when the 
colliding wave packets of different color start overlapping,
we observe an increasing longitudinal energy density in
the overlap region around the center of collision at
$x_3=100\,{\rm fm}$. Fig. 2 clearly shows that $w_L^{(E)}(t,z)$
grows rather fast from time step $t_{1000}$ until time step
$t_{1500}$ where it reaches a maximum and has grown
by more than two orders of magnitude.
For larger times the hump decreases again and practically
disappears at $t_{3000}$. Around $t_{6000}$, however, the longitudinal 
energy density grows again at the positions of the receding
wave packets. After 10000 time steps $w_L^{(E)}(z)$ has
increased by five orders of magnitude.

A comparison of Fig. 1 with Fig. 2 indicates that the total
color electric field energy on transverse links of the lattice
decreases at large times while the total
color electric field energy on longitudinal links
increases. In order to understand this behavior in detail, 
we explore the time dependence of these quantities for
different values of the coupling constant. 
To be more precise,
we define the transverse and longitudinal energy of the color 
electric fields as
\begin{eqnarray}
\label{Eq.26a}
W_T^{(E)}(t) &=&\int d^3x \sum\limits_{l=1}^2
      {\rm Tr}\bigl({\cal E}_l({t,\vec x}){\cal E}_l{({t,\vec x})}\bigr),
\\
\label{Eq.26b}
W_L^{(E)}(t) &=& \int d^3x
       {\rm Tr}\bigl({\cal E}_3{({t,\vec x})}{\cal E}_3{({t,\vec x})}\bigr)
\end{eqnarray}
The integration in the definitions (\ref{Eq.26a}) and (\ref{Eq.26b})
is carried out over the whole lattice.

Fig. 3 displays $W_T^{(E)}(t)$ over a time interval of $600a$
which corresponds to 6000 time steps. 
The energies are compared for different values of the coupling constant $g$.  
As already mentioned above, 
values of the coupling constant which differ from $g=1$ can be
scaled out with the amplitude of the wave functions. The amplitude of
wave packets, however, determines the energy carried by a wave.
Once the energy is fixed by describing a real physical system, it
makes sense to use different values of the coupling constant. 
Here, we collide wave packets which are normalized
to ${N_1N_2a^2}/\sigma$ particles and refer to different values of $g$. 
%
%
%
%
%
%
\begin{figure}[H]
\centerline{
\epsfysize=14cm \epsfxsize=14cm
\epsffile{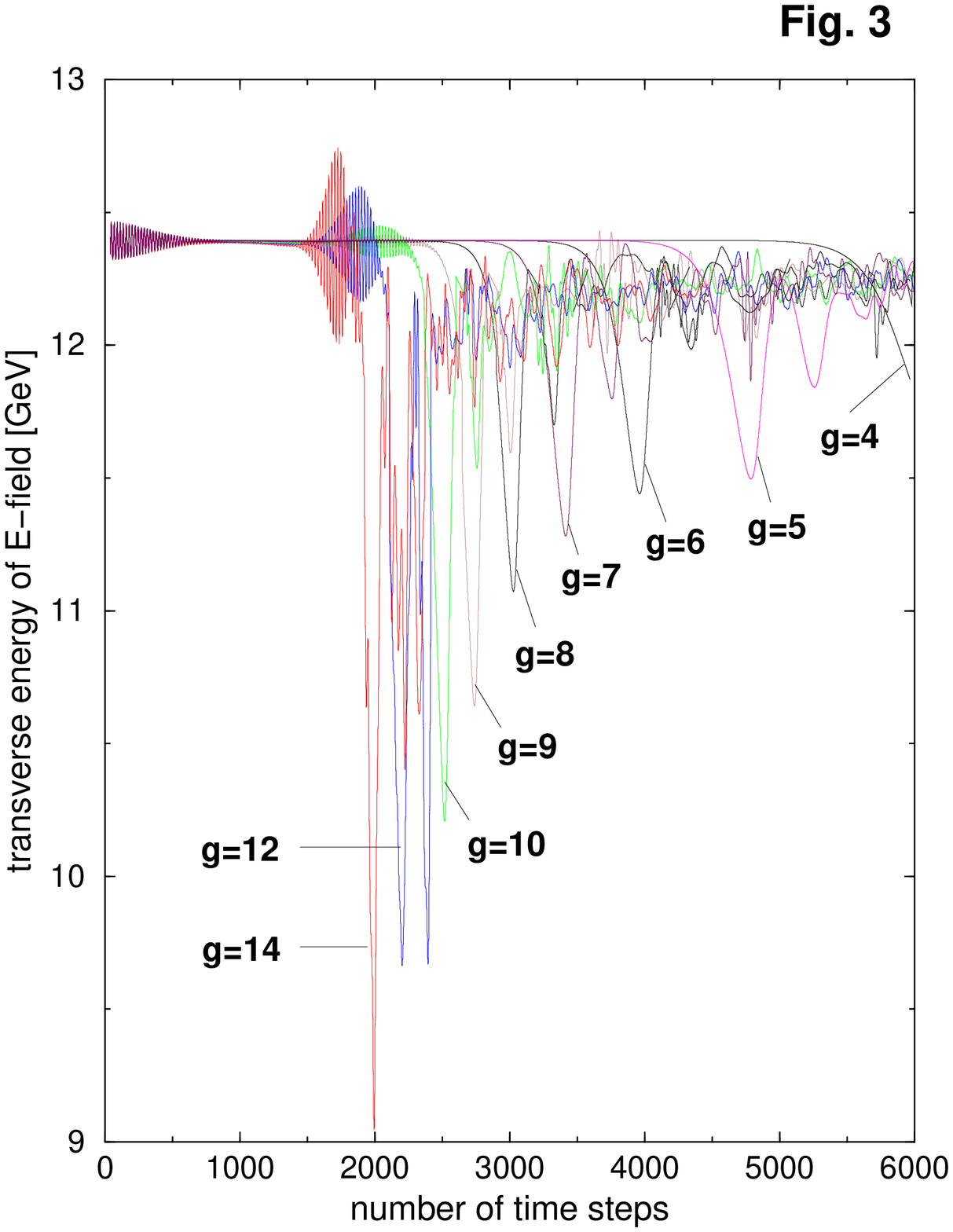}
}
\caption{The transverse color electric field energy $W_T^{(E)}(t)$ as
         a function of time is displayed for different values of the
         coupling constant $g$. The time step width of each time step
         is $\Delta t = a_3/10 = 0.01\,{\rm fm}$. }
\end{figure}
Fig. 3 shows that for small coupling, $W_T^{(E)}(t)$
remains unchanged through a long period of time after the collision
which occurs in an interval around the time step $t_{1500}$. 
For $g=4$ the transverse color electric energy starts to
decrease around time step $t_{5000}$. A comparison with curves
obtained for increasing values of $g$ shows that the $W_T^{(E)}(t)$ begins
to decrease at earlier times. For the largest values $g=10,12,14$,
the decrease begins in the overlap region of the wave packets. 
In these cases, strong oscillations occur in the overlap region.
The magnetic field energies $W_T^{(B)}(t)$ show a very similar behavior.
The total energy of both wave packets is $24.76\,{\rm GeV}$
for our numerical choice $a= 0.1\,{\rm fm}$.

Fig. 4 displays the corresponding longitudinal field energy
$W_L^{(E)}(t)$. Before the wave packets start to overlap 
we find $W_L^{(E)}(t)=0$ to a very high precision ($10^{-26}$).
In the overlap region, a hump like structure occurs which grows
for increasing values of $g$. For small values of $g$, $W_L^{(E)}(t)$
vanishes after the wave packets have passed through each other. 
After a long time $W_L^{(E)}(t)$ grows very steeply. Fig. 4 shows
clearly that the time between overlap and sudden growth shrinks
for increasing coupling $g$. We find, that this time difference
scales like $1/g^2$.
%
%
%
%
%
%
\begin{figure}[H]
\centerline{
\epsfysize=14cm \epsfxsize=14cm
\epsffile{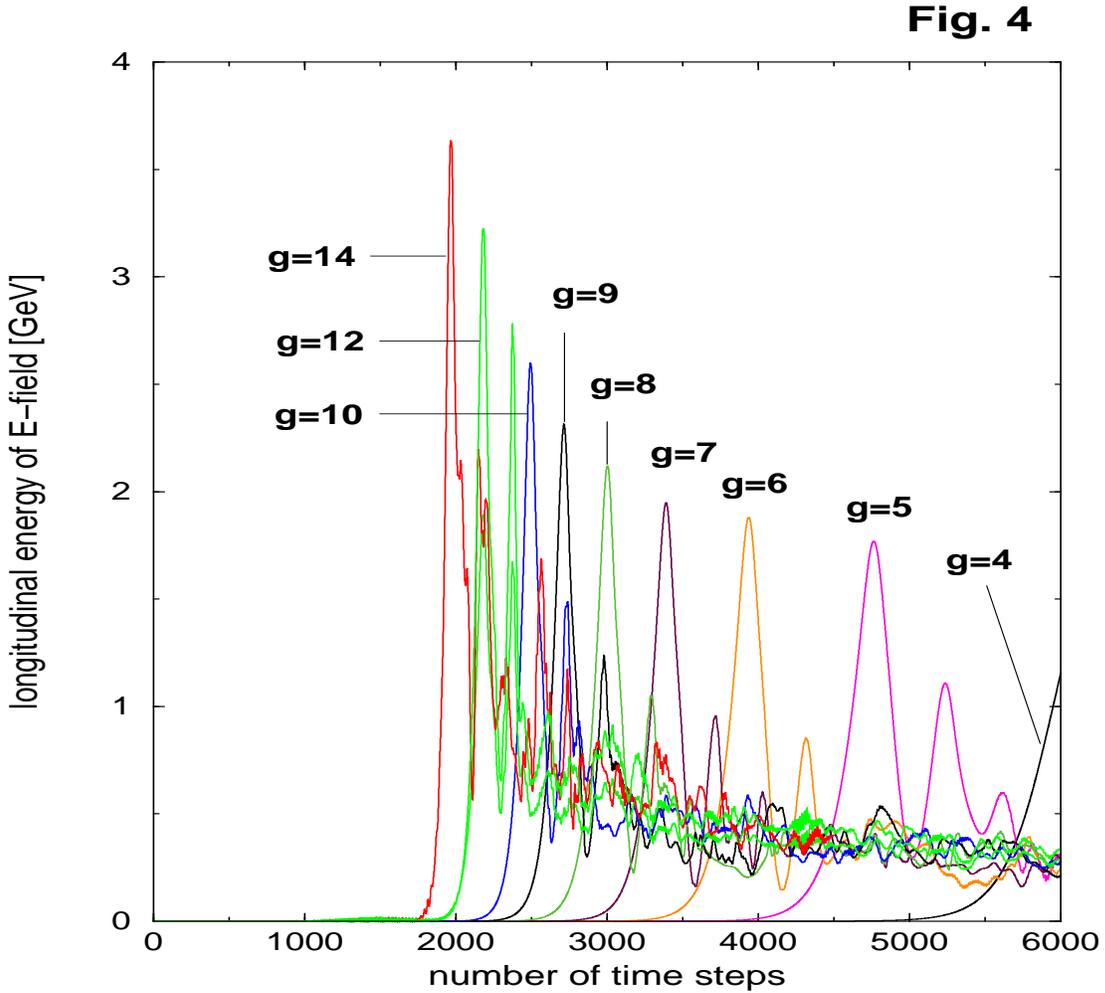}
}
\caption{The longitudinal color electric field energy $W_L^{(E)}(t)$ as
         a function of time is displayed for different values of the
         coupling constant $g$. The time step width of each time step
         is $\Delta t = a_3/10 = 0.01\,{\rm fm}$. }
\end{figure}
Now the question arises whether the
energy deposit in longitudinal links can be associated with
fields propagating into transverse directions.
To define energy currents, we denote the Poynting vector in the
adjoint representation
\begin{equation}
\label{Eq.27}
\vec {\cal S} := c\, \vec {\cal E}\times {\cal B}.
\end{equation}
With the transverse and longitudinal components of the vector 
(\ref{Eq.27}) we define the total transverse and longitudinal
energy currents
\begin{eqnarray}
\label{Eq.28a}
i_T(t):= \sum_{l=1}^2 \int d^3x 
          2\,\big\vert{\rm Tr}
          \bigl( {\cal S}_l(t,\vec x) \bigr)
          \big\vert
\\
\label{Eq.28b}
i_L(t):= \sum_{l=3}^3 \int dx^3 
          2\,\big\vert{\rm Tr}
          \bigl( 
          {\cal S}_l(t,\vec x)
          \bigr)
          \big\vert .
\end{eqnarray}
In the following Fig. 5, we show the total transverse energy current 
$i_T(t)$ in the overlap time region for different values of the coupling 
constant. The figure shows that there is no transverse energy current
before the wave packets start to overlap. In the overlap time
region between the time steps $t_{800}$ and $t_{1500}$, we observe 
a strong increase of the transverse energy current. The maximum
is reached at time step $t_{1500}$ where the overlap is maximal.
At decreasing overlap in time region between $t_{1500}$ and $t_{2200}$,
$i_T(t)$ vanishes. The height of the humps increases for increasing
values of $g$ and scales like $1/g^2$. To become more precise, in the 
time region of the humps as shown in Fig. 5, the transverse current 
implies an oscillating structure
of the time period $T_t = \overline\lambda/4$ over which we have averaged 
to obtain the smooth curves in Fig. 5. $\overline\lambda$ is here
defined as $\overline\lambda = {(2\pi)/k_3}$ 
In the upper left panel in Fig. 5, we
show an example for $\overline k_3= \pi/(4a_3)$ in the case $g=8$
in which no averaging has been done. The curve displays a period
of $T_t=2a_3$ according to $\overline\lambda = 8a_3$. No averaging
was necessary for times after the overlap time region.      
%
%
%
%
%
\begin{figure}[H]
\centerline{
\epsfysize=14cm \epsfxsize=14cm
\epsffile{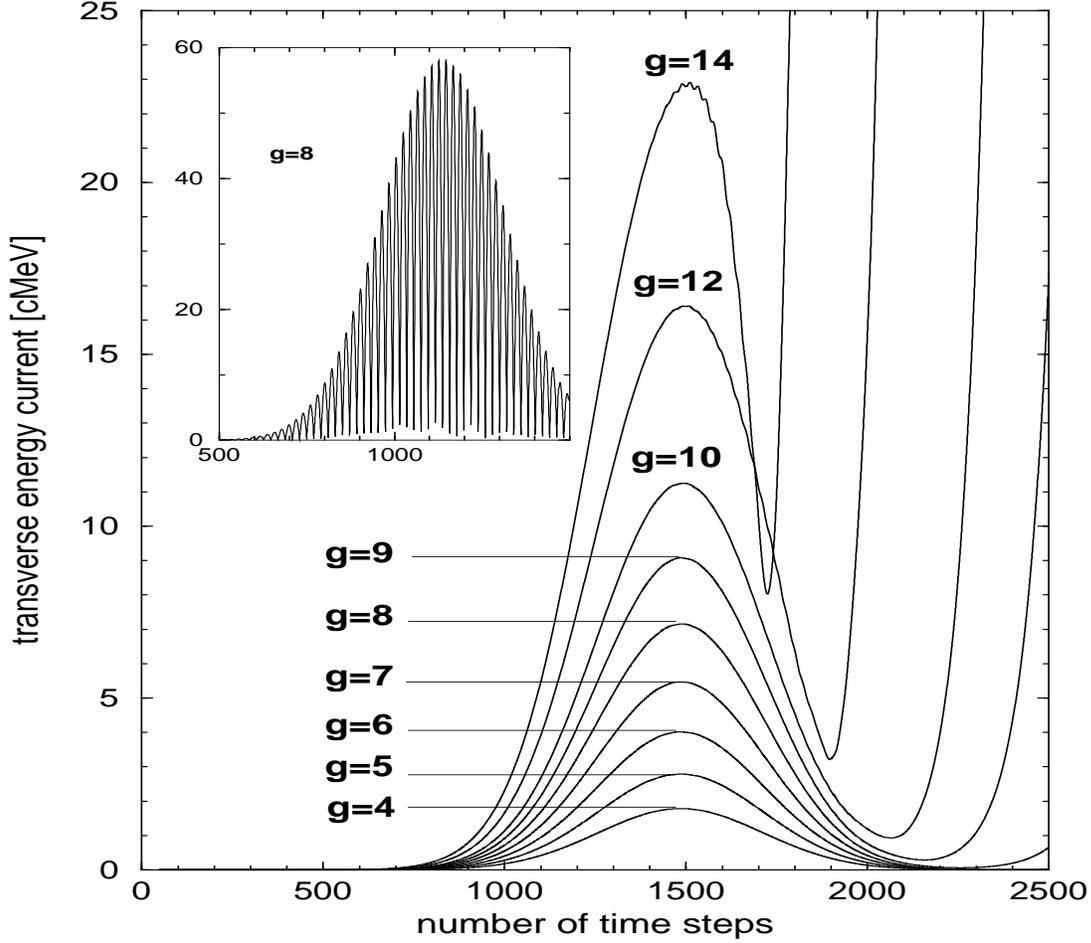}
}
\caption{The total transverse energy current $i_T(t)$ as
         a function of time is displayed for different values of the
         coupling constant g. The average longitudinal momentum of the 
         colliding wave packets is $\overline k_3=\pi /(2a_3)$. 
         In the upper left window, the full time-dependence of the transverse
         energy current is displayed at the example $g=8$ and 
         $\overline k_3=\pi /(4a_3)$. } 
\end{figure}
Fig. 6 displays the same transverse energy currents as Fig. 5 
but for a larger time interval and on a larger scale. Note that
the current is plotted in units of $[i_T(t)]=c\,{\rm GeV}$ in Fig. 6
but in units $[i_T(t)]=c\,{\rm MeV}$ in Fig. 5.
After passing through the hump region between time step $t_{800}$
and $t_{2000}$, the energy current $t_T(t)$ disappears. For the
smallest coupling $g=4$, it takes about 3500 time steps
or $\tau\simeq 35\,{\rm fm}$ from maximum overlap until to the point
where $i_T$ 
starts to re-grow. This time, $i_T$ grows to much larger values
and stays large at later times. The time delay $\tau$ depends on 
$g,\Delta k_3,\overline k_3$. We find that $\tau \sim 1/g^2 $ but
haven't studied carefully the dependence on $\Delta k_3$ and 
$\overline k_3$. For large $\overline k_3$ at least, one can 
argue analytically that $\tau\sim {\overline k_3}^{-3/4}$. This
will be further discussed in the appendix B.
Some few calculations for smaller $\overline k_3$ have shown
that $\tau$ increases monotonically for decreasing $\overline k_3$. 
The sudden increase of $i_T$ defines the ``glue burst''. For large
coupling it occurs already in the overlap region.
%
%
%
%
%
\begin{figure}[H]
\centerline{
\epsfysize=14cm \epsfxsize=14cm
\epsffile{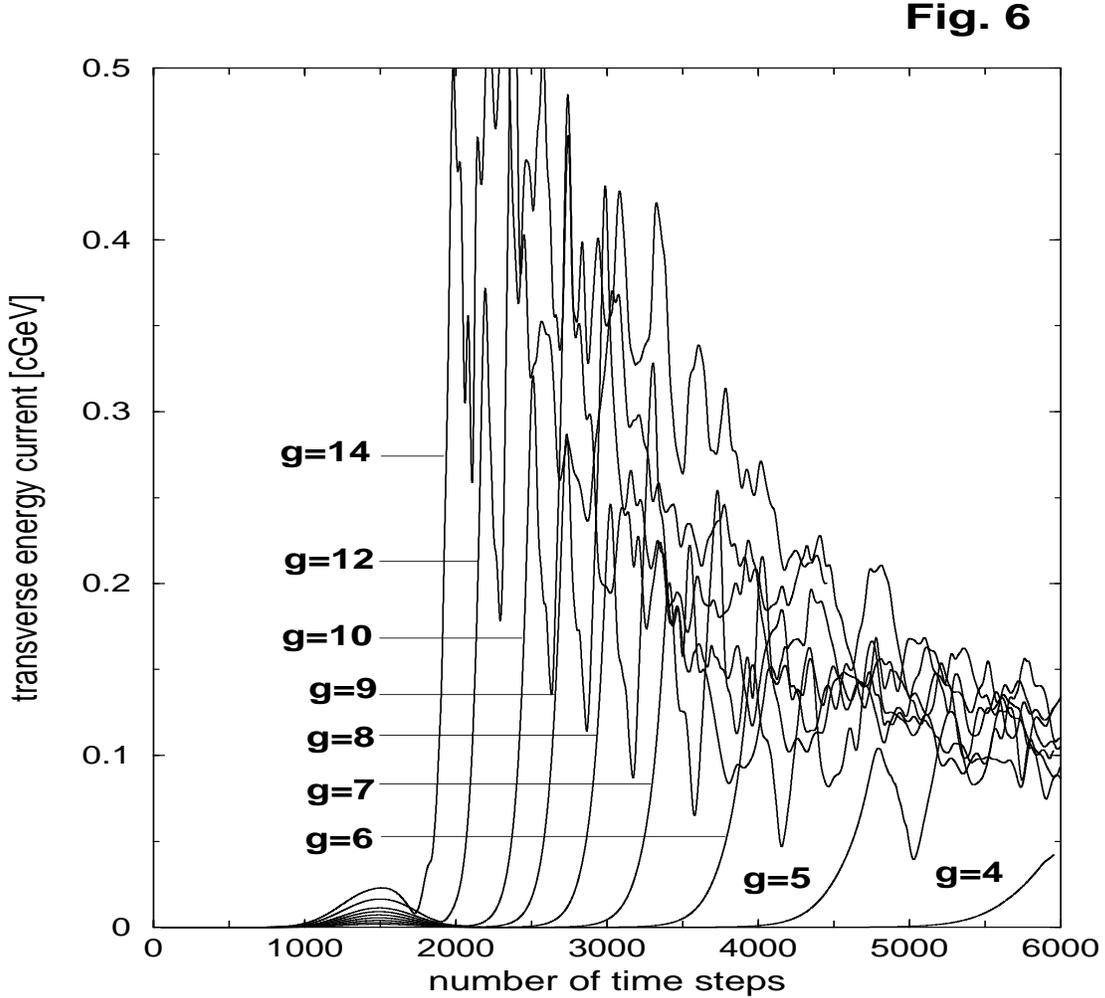}
}
\caption{ Same as in Fig. 5 but for a larger time intervall
          and on a larger scale for the transverse energy
          current $i_T(t)$. 
}
\end{figure}
The following Fig. 7 displays the corresponding longitudinal
energy currents $i_L(t)$. It shows that $i_L$ decreases
at the same time when $i_T$ increases. 
%
%
%
%
%
\begin{figure}[H]
\centerline{
\epsfysize=14cm \epsfxsize=14cm
\epsffile{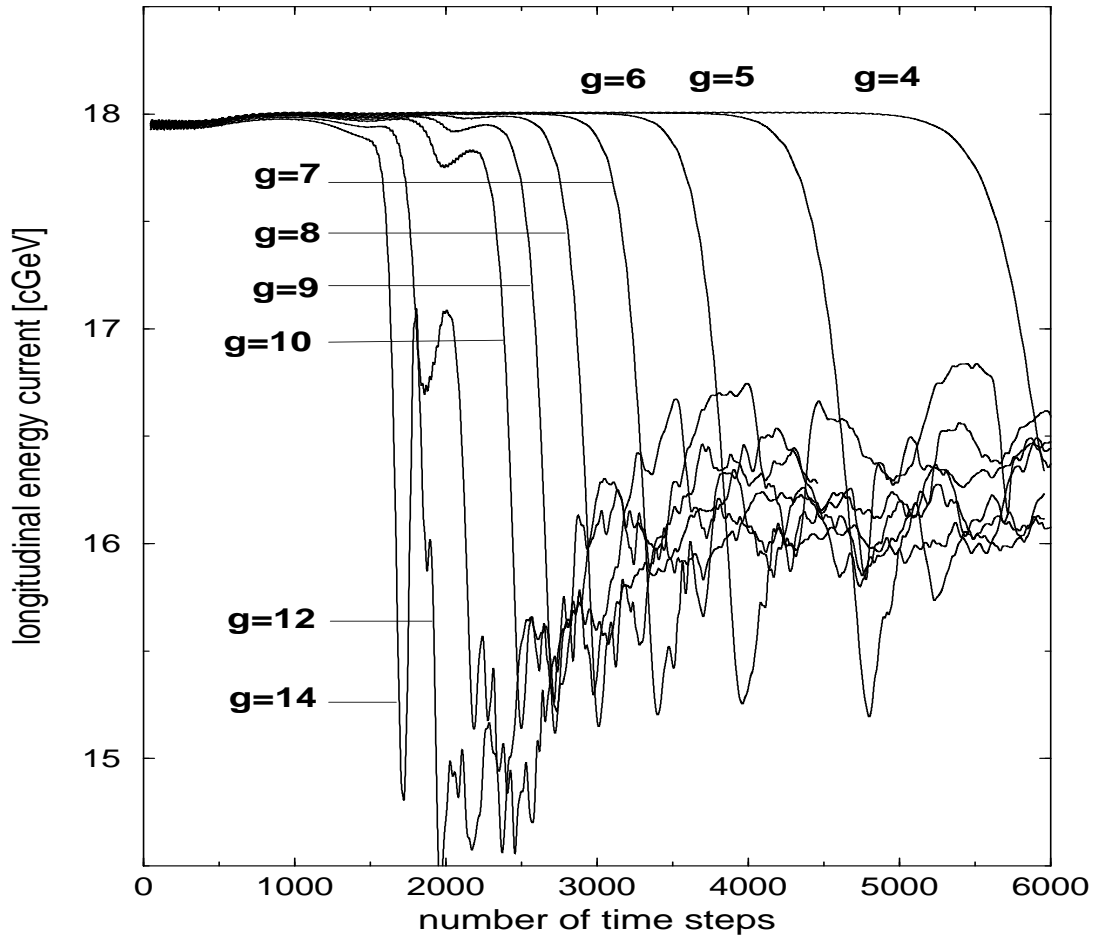}
}
\caption{ The longitudinal energy currents $i_L(t)$ corresponding
          to the cases shown in Fig. 6 are displayed.         
          $i_L(t)$ decreases at the same time when $i_T(t)$
          increases.
}
\end{figure}
Subsequently, we present some results from the collision of
wave packets with a finite transverse extent. 
We have carried out similar calculations
as in the case of plane waves and found a similar, but
even more pronounced behavior
for the time dependence of the transverse and  
longitudinal energy currents.
Most of the calculations were carried out on
a lattice of the size 4x200x800 points.
The wave packets have been initialized in the same
way as in the case for plane wave packets.
However, the initial wave functions depend now also on the 
transverse coordinates.
Since a lattice with the same extension in both transverse directions
would lead to an exceedingly large number of lattice points, we chose
a small number of lattice points in the $x_1$-direction.
Thus, we study the dynamics of a collision in the $(x_2,x_3)$-plane
integrated over the $x_1$-coordinate.
Under these restrictions, 
the scalar function $\phi(t,x_2,x_3)$ defines the initial wave packet
\begin{eqnarray}
\label{Eq.29}
\phi(t,x_2,x_3)&:=&\phi_0
\exp{\bigl(-{1\over 2}\Delta k_3^2(t+x_3)^2\bigr)}
\exp{\bigl(-(\Delta k_2x_2)^{\alpha}\bigr)}
\cos{\bigl(\overline k_3(t+x_3)\bigr)}, \\
\label{Eq.30}
\phi_0 &:=&
\sqrt{ {(2\Delta k_3)}/({\sqrt{\pi}\sigma\overline k_3})}. 
\end{eqnarray}
The additional factor $\exp{\bigl(-(\Delta k_2 x_2)^{\alpha}\bigr)}$
describes the shape of the wave packet in the transverse $x_2$-direction.
The parameter $\Delta k_2$ determines the transverse width and the
parameter $\alpha$ controls the transverse shape. For large values
of $\alpha$, we obtain a sharp surface and the exponential tails
are suppressed. Since the transverse extension of the lattice is limited,
we choose the extension of the initial wave packet in the 
$x_2$-direction small enough to leave space for the transverse dynamics
after the collision. Good values are $\Delta k_2 = 1/(40 a_2)$ and
$\alpha = 8$. The wave packet is thus $8\,{\rm fm}$ wide and extends
over 80 lattice points into the main transverse direction. Further outside
the amplitude is practically zero since $\alpha $ is large. 
For $\Delta k_3 = \pi/(100a_3)$, the FWHM in longitudinal direction
is $10\,{\rm fm}$.

With the function $\phi(t,x_2,x_3)$
we determine the initial link variables and the color electric
field amplitudes and map these data on the lattice at the initial 
positions of the wave packets. For a collision, the wave packets
have opposite average momenta $\pm\overline k_3$. 
The evolution of the collision is carried out in the same way
as for plane wave packets. 
Fig. 8, shows snapshots of the transverse energy density of the color 
electric fields which we define as 
\begin{eqnarray}
\label{Eq.31}
w_T^{(E)}(t,x_2,x_3) &=& \int dx_1 \sum\limits_{l=1}^2
                         {\rm Tr}\bigl({\cal E}_{l}(t,\vec x)
                         {\cal E}_{l}(t,\vec x) \bigr).
\end{eqnarray}
The simulation shown in Fig. 8 has been carried out 
for the non-Abelian case and 
with a coupling $g=6$. 
The other parameter settings are as for plane waves above.  
The upper left picture shows the distribution 
for the time step $t_{100}$ where the shape has almost not
changed as compared to the initial distribution. The wave packets
decay into the transverse directions while they propagate free over
the lattice. The upper right picture displays 
$w_T^{(E)}(t_{1500},x_2,x_3)$ at maximum overlap.
In the middle left $w_T^{(E)}(t_{3500},x_2,x_3)$ is displayed.
The distribution is almost identical with that obtained in the Abelian case
at the same time step $t_{3500}$. At the time step $t_{4000}$,
however, the waves packets start to burst. 
The last two images correspond to the time steps
$t_{5000}$ and $t_{6000}$.
%
%
%
%
%
\begin{figure}[H]
\centerline{
\epsfysize=14cm \epsfxsize=14cm
}
\caption{ The transverse color electric field energy density
          $w_T^{(E)}(t,x_2,x_3)$ is shown for six selected
          snapshots taken at the time steps $t_{100}$, 
          $t_{1500}$, $t_{3500}$, $t_{4000}$, $t_{5000}$
          and $t_{6000}$. The corresponding pictures
          are ordered from the upper left to the lower right.
}
\end{figure}

\noindent
Note: The Fig.8 exceeds the size accepted at the lanl-server and
was therefore omitted. We recommend to download the postscript file
from the web page http://www.phy.duke.edu/\~\,poeschl/ under 
``Rsearch Related Links'' and ``Preprints''.
%
%
%
%
%
%
%
%
%
\section {Analysis of the Yang Mills Equations}
%
%
In the following, we present the explanation of the glue burst
solution by analyzing the Yang Mills equations (\ref{Eq.1}).
The details of the discussion are presented in the appendix A.
With the definition of the field tensor 
${\cal F}^{\mu\nu}:={\cal D}^{\mu}{\cal A}^{\nu} -
{\cal D}^{\nu}{\cal A}^{\mu}$ , we rewrite Eq.
(\ref{Eq.1}) for the gauge fields in the adjoint denotation
\begin{equation}
\label{Eq.40}
\bigl[{\cal D}_{\mu},\bigl( {\cal D}^{\mu}{\cal A}^{\nu} -
{\cal D}^{\nu}{\cal A}^{\mu}\bigr)\bigr]_- = 0
\end{equation} 
With the definition of the derivative ${\cal D}^{\mu}$ in
Eq. (\ref{Eq.2}) follows
\begin{eqnarray}
\label{Eq.41}
\partial_{\mu}\partial^{\mu}{\cal A}^{\nu}&=&
\partial_{\mu}\partial^{\nu}{\cal A}^{\mu} \,+\, ig\partial_{\mu}
\bigl[ {\cal A}^{\mu},{\cal A}^{\nu}\bigr]_-  
\nonumber \\
& + &ig\bigl[{\cal A}_{\mu},\partial^{\mu}{\cal A}^{\nu}\bigr]_-
\nonumber \\
& - &ig\bigl[{\cal A}_{\mu},\partial^{\nu}{\cal A}^{\mu}\bigr]_- 
\nonumber \\
& + &g^2\bigl[{\cal A}_{\mu},
        \bigl[{\cal A}^{\mu},{\cal A}^{\nu}\bigr]_-\bigr]_-  
\end{eqnarray}
Since the time delayed burst-like behavior of the solution
occurs also in the case of colliding plane wave packets, 
we first consider this case. 
The phenomenon occurs in two steps. 
It is based on a delicate interplay of
color degrees of freedom between the color
sub-spaces ${\rm span}(T_1,T_2)$ and ${\rm span}(T_3)$
and between transverse and longitudinal degrees of freedom
in the field amplitudes. 

The first step occurs in the overlap region of the
wave packets in both, time and space.
The Yang Mills equations provide a mechanism which transfers
field energy from transverse into longitudinal field degrees
of freedom. It is explained by the equation for the 
longitudinal gauge field components which follows from
Eq. (\ref{Eq.41}) for ${\nu = 3}$. For plane wave packets,
only the fourth term and the $g^2$-term do not vanish 
on the r.h.s. of Eq. (\ref{Eq.41}) in the case $\nu = 3$
and therefore the equation
\begin{eqnarray}
\label{Eq.42}
\partial_{\mu}\partial^{\mu}{\cal A}^{3}&=&
 \,-\, ig\bigl[{\cal A}_{\mu},\partial^{\nu}{\cal A}^{\mu}\bigr]_- 
\nonumber \\
&  &\,+\,g^2\bigl[{\cal A}_{\mu},
        \bigl[{\cal A}^{\mu},{\cal A}^{\nu}\bigr]_-\bigr]_-  
\end{eqnarray} 
describes the dynamics of the longitudinal fields
${\cal A}^3$ during the overlap time.
Assuming that 
$\big\vert A_{\mu}^3  A^{\mu}_3\big\vert \ll
  \big\vert A_{\mu}^1  A^{\mu}_1\big\vert $
and  
$\big\vert A_{\mu}^3  A^{\mu}_3\big\vert \ll
 \big\vert A_{\mu}^2  A^{\mu}_2\big\vert $ during the 
overlap time, we find the following leading
contributions on the r.h.s. of Eq. (\ref{Eq.42}).   
\begin{eqnarray}
\label{Eq.43}
-ig\bigl[{\cal A}_{\mu},\partial^3{\cal A}^{\mu}\bigr]_- 
& \simeq &
{g}\bigl( A_{\mu}^1\partial^3 A_2^{\mu} -
                 A_{\mu}^2\partial^3 A_1^{\mu}\bigr) T_3
\\
\label{Eq.44}
g^2\bigl[{\cal A}_{(1)},
    \bigl[{\cal A}^{(1)},{\cal A}^{(3)}\bigr]_-\bigr]_-  
& = &
\,{{g^2}} A_{(1)}^1A_1^{(1)}A_3^{(3)}\, T^3 
\,-\,{{g^2}} A_{(1)}^3A_1^{(1)}A_3^{(3)}\, T^1 
\nonumber \\
&  &
\,+\,{{g^2}} A_{(1)}^2A_2^{(1)}A_3^{(3)}\, T^3 
\,-\,{{g^2}} A_{(1)}^3A_2^{(1)}A_3^{(3)}\, T^2 
\end{eqnarray}
The above assumption is certainly true for the numerical results 
presented in the previous section which have demonstrated that 
in the overlap time the
longitudinal energy densities are smaller by at least one
order of magnitude in comparison to the transverse energy
densities. The indices of the space coordinates in 
Eq. (\ref{Eq.44}) are now denoted in parentheses in order to
distinguish from color indices.
In leading approximation, the dynamics of the longitudinal 
fields in the overlap time is thus described by the following 
equation
\begin{eqnarray}
\label{Eq.45}
\partial_{0}\partial^{0}{\cal A}^{3} & \simeq &
{g}\bigl( A_{\mu}^1\partial^3 A_2^{\mu} -
                 A_{\mu}^2\partial^3 A_1^{\mu}\bigr) T_3 
\nonumber \\
&  &
\,+\,{{g^2}} A_{(1)}^1A_1^{(1)}A_3^{(3)}\, T^3 
\,-\,{{g^2}} A_{(1)}^3A_1^{(1)}A_3^{(3)}\, T^1 
\nonumber \\
&  &
\,+\,{{g^2}} A_{(1)}^2A_2^{(1)}A_3^{(3)}\, T^3 
\,-\,{{g^2}} A_{(1)}^3A_2^{(1)}A_3^{(3)}\, T^2. 
\end{eqnarray}   
The order $g$ term on the r.h.s. of Eq. (\ref{Eq.45}) 
acts as a source term for the third color component of the 
longitudinal gauge field ${\cal A}^{3}(t,\vec x)$.
As soon as the wave packets start to penetrate into each other,
the space integral over this source term 
\begin{equation}
\label{Eq.46}
I_3(t):= \int d^3 x {g}\bigl( A_{\mu}^1(t,\vec x)
         \partial^3 A_2^{\mu}(t,{\vec x}) -
         A_{\mu}^2(t,{\vec x})\partial^3 A_1^{\mu}(t,\vec x)\bigr)
\end{equation}
starts to grow. As Fig. 9 shows
in a simplified manner, the derivative
$\partial^3 A_1^{\mu}$ is negative in the overlap
region at the beginning and $\partial^3 A_2^{\mu}$
is positive.   
Consequently, $I_3(t)>0$ at times before the wave packets
have reached maximum overlap. After the maximal overlap
the derivatives change the sign and $I_3(t)<0$. 
%
%
%
%
%
\begin{figure}[H]
\centerline{
\epsfysize=14cm \epsfxsize=14cm
\epsffile{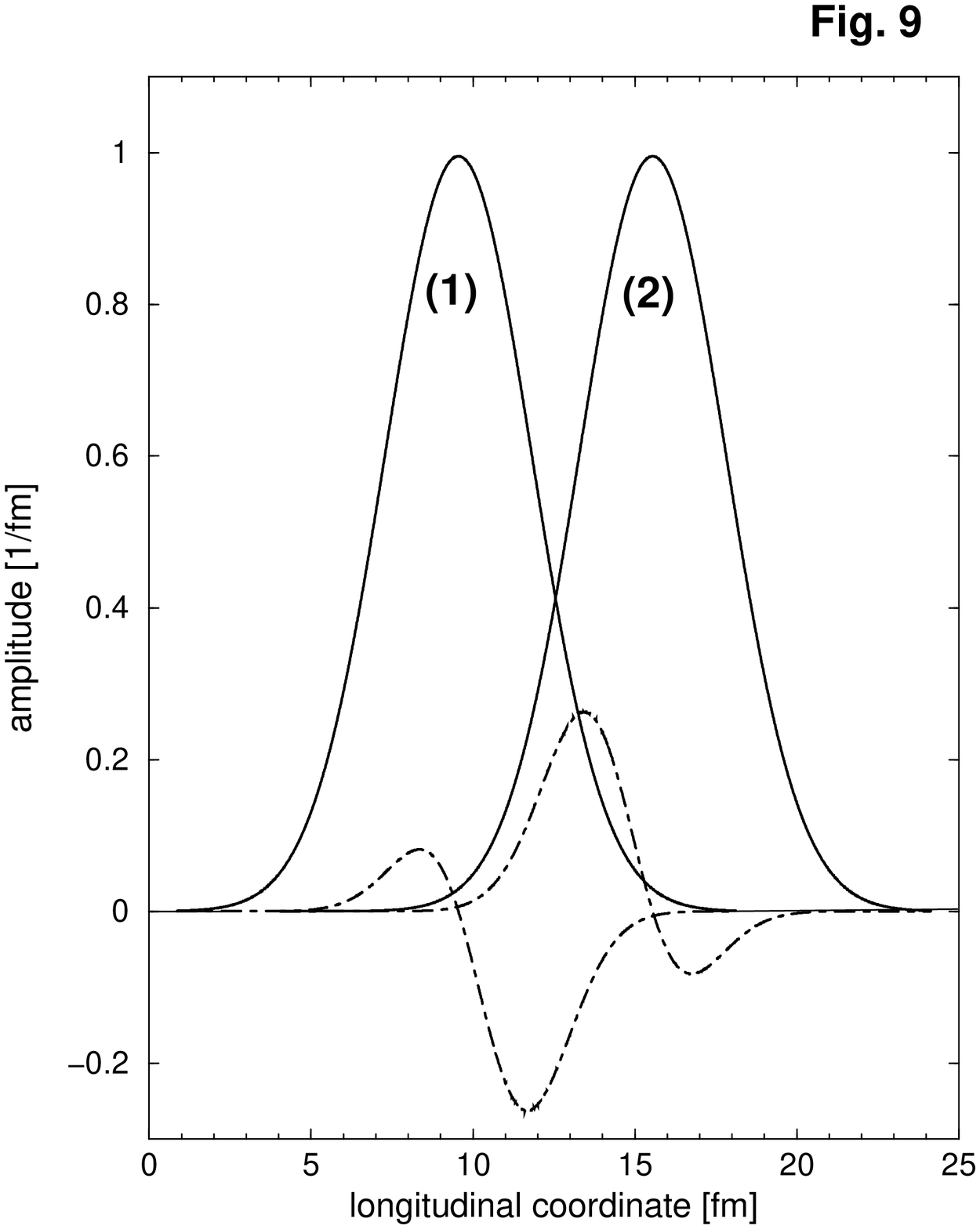}
}
\caption{ Two colliding Gaussian wave packets of gauge fields $A_{1}$ and 
          $A_{2}$ for which 
          $\phi(t,x_3)=\phi_0\exp(-1/2\Delta k_3^2 (t+x_3)^2)$  
          at the beginning of the overlap. The products
          $A_{\mu}^1\partial^3 A_2^{\mu}$ and $A_{\mu}^2\partial^3 A_1^{\mu}$ 
          (plotted as dashed lines) of the field amplitudes (solid lines)
          contribute to a source current in the longitudinal Yang-Mills
          equation. }              
\end{figure}
The Fig. 9 displays in a qualitative manner the situation for the most simple
case when $\overline k_3 =0$ and the wave packets 
show no oscillation in position space. 
The situation shown in Fig. 9 in principle occurs for each pair of
overlapping humps in colliding oscillating wave packets for which
$\overline k_3 >0$. 
The corresponding time dependent behavior of the source current 
$I_3(t)$ at $\overline k_3 =0$  
is depicted in Fig. 10. When the penetrating wave packets
increase their overlap, the longitudinal contribution $I_3(t)$ to the source 
increases first. Since the derivative on the opposite side of both
wave packets has opposite sign, the source term changes the sign
as soon as the two humps have passed the maximum overlap.
For $\overline k_3 >0$, the wave packets depicted in
Fig. 9 would contain oscillations, i.e. 
the number of oscillations of the curves in Fig. 10 
and in the source term would increase accordingly. In this
case the envelope of the curves would be of gaussian shape
rather than the curves itself.
%
%
%
%
%
\begin{figure}[H]
\centerline{
\epsfysize=14cm \epsfxsize=14cm
\epsffile{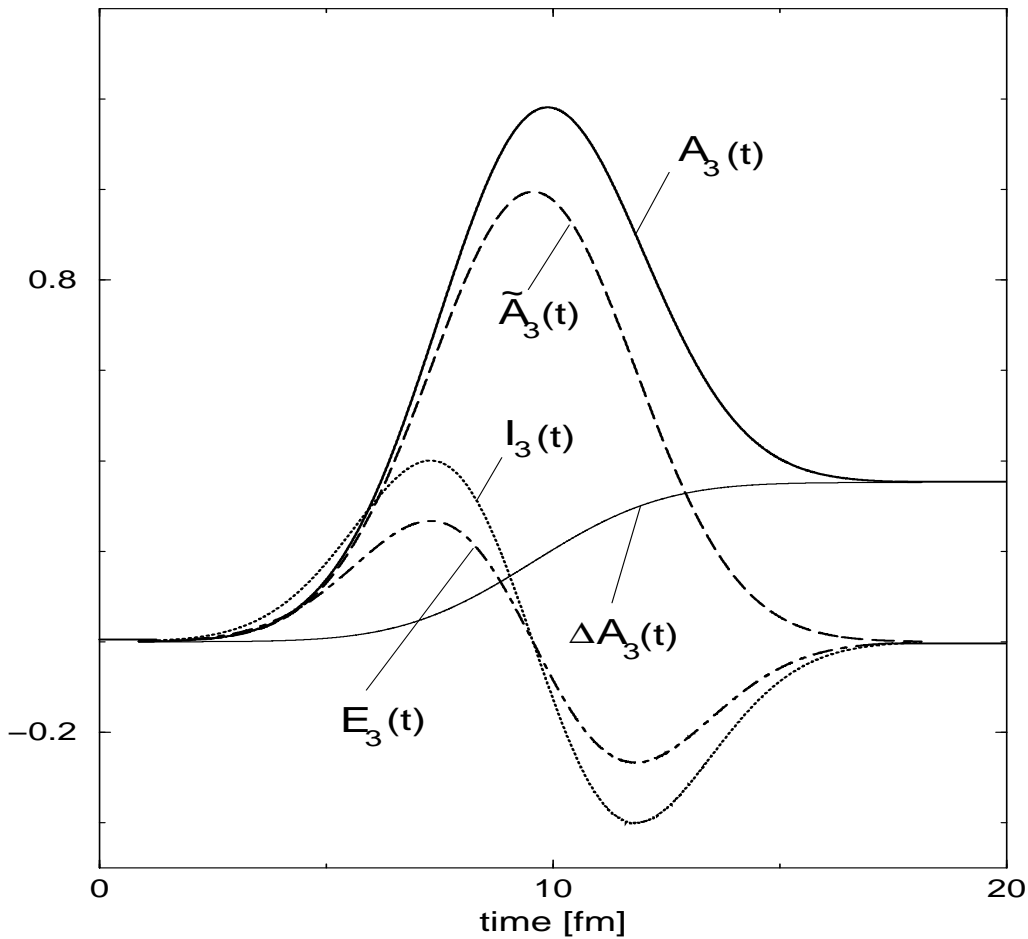}
}
\caption{ The longitudinal contribution $I_3(t)$ to the source
          is plotted as a function of time (doted line).
          The time dependence of the source term leads to a
          time dependence of the gauge field ${\cal A}_3$ as indicated
          by the dashed curve. The corresponding color electric
          field is indicated by the dot-dashed curve. The
          contribution from the $g^2$-term $\Delta {\cal A}_3(t)$
          (solid line)
          adds to the gauge field resulting in the total field
          ${\cal A}_3(t)$ (fat solid line). 
         }              
\end{figure}
As the source term     
${g}\bigl( A_{\mu}^1\partial^3 A_2^{\mu} -
                  A_{\mu}^2\partial^3 A_1^{\mu}\bigr)$
grows in time, the longitudinal gauge field ${\cal A}^3(t,\vec x)$ 
grows. At the time of maximal overlap the source current
changes the sign and cancels the gauge field $A^3(t,\vec x)$ 
as shown qualitatively by the dashed curve in Fig. 10.

During the whole time interval in which the wave packets
pass through each other the finite longitudinal gauge
field ${\cal A}^3(t,\vec x)$ enters into the last
term on the r.h.s. of Eq. (\ref{Eq.41}) leading to
an additional but small contribution to the source term
as long as $g$ is not too large.
For large $g^2$, the $g^2$-term takes 
over during the overlap time as soon as the order $g$ term
has generated a finite amplitude $\tilde {\cal A}^3$. 

Our calculations show that for not to large values of $g$ 
($g<10$) and for small overlap times the contribution 
from the $g^2$-term is not dominant compared to the contribution 
resulting from the first source term on the r.h.s of Eq. (\ref{Eq.45}). 
Furthermore, the $g^2$-term can not act as a source as long as
${\cal A}^3=0$. 
The finite longitudinal amplitude
has first to be generated by the fourth term 
on the r.h.s. of Eq. (\ref{Eq.41}) or the first
term on the r.h.s of Eq. (\ref{Eq.45}) respectively 
which acts as a initiator for the $g^2$-term. 
The fourth term also again switches off its 
contribution to the longitudinal gauge field 
when the wave packets have passed the maximum overlap.
The fourth term goes to zero at vanishing overlap
and doesn't contribute any further in the receding wave
packets. 
However, the $g^2$ term causes a finite but small contribution
$\Delta {\cal A}^3(t)$ 
to the longitudinal gauge field during the overlap time.
The qualitative time dependent behavior of this contribution is shown
by the solid line in Fig. 10.
$\Delta {\cal A}^3(t)$ which is not canceled after the overlap is
proportional to the surface under the dashed curve 
($\tilde {\cal A}_3(t)$) in Fig. 10. The fat solid curve depicts
the time dependent behavior of the total longitudinal field ${\cal A}^3(t)$
while the dashed curve is obtained without $g^2$-term. 

As a consequence of this mechanism, a small
longitudinal color field component 
${\cal A}^3(t,\vec x)$ is left over in the 
receding wave packets. This contribution initiates
the second step which forms the burst itself. 
Without restriction of the general case, we assume
that the first wave packet is initially polarized in ${\rm span}(T_1)$
and the second wave packet is polarized in ${\rm span}(T_2)$.
As soon as the wave packets obtain a small contribution
in the gauge field component $A_3^{(3)}$, the spacial component
${\cal A}^{(1)}$ starts to change the polarization in color space.
In wave packet (1) the components $A_2^{(1)}$ and $A_3^{(1)}$ 
grow whereas $A_1^{(1)}$ decreases accordingly.    
In wave packet (2) the components $A_1^{(1)}$ and $A_3^{(1)}$ 
grow whereas $A_2^{(1)}$ decreases accordingly.    
This behavior results in a rapid change of the $T^3$-terms 
on the r.h.s. of Eq. (\ref{Eq.45}).
It is explained by the Yang-Mills equation for
the first component which follows from Eq. (\ref{Eq.41})
for $\nu = 1$. Analyzing the r.h.s. terms, we find
\begin{eqnarray}
\label{Eq.47}
\partial_{\mu}\partial^{\mu}{\cal A}^{1} & = &
\,-\,g A_1^{(1)}\partial_{(3)} A_3^{(3)} T^2
\,+\, g A_1^{(1)}\partial_{(3)} A_3^{(3)} T^1
\nonumber\\
&  &
\,-\, g A_3^{(3)}\partial_{(3)} A_1^{(1)} T^2
\,+\, g A_3^{(3)}\partial_{(3)} A_2^{(1)} T^1 \,+\, ....
\nonumber\\
&    &\, +\, g^2 A_3^{(3)}A_3^{(3)}A_1^{(1)} T^1
\,+\, g^2 A_3^{(3)}A_3^{(3)}A_2^{(1)} T^2 \,+\, ....  
\end{eqnarray}
where we denote only the most important contributions
on the r.h.s.\,. The first four terms on the r.h.s. come from the
second term on the r.h.s. of Eq. (\ref{Eq.41}).
Similar contributions which come from the third term are
omitted for brevity since they do not bring a new
aspect into the discussion.
The fourth term does not contribute for plane waves.
The last two terms on the r.h.s. of Eq. (\ref{Eq.47}) 
come from the $g^2$-term. For details, we refer to
the appendix A.

In the following we discuss the mechanism described by
Eq. (\ref{Eq.47}). 
The first, third and sixth term on the r.h.s. lead to excitations
of modes in the color direction $T^2$ in the wave packet
(1). The second, fourth and fifth term act in a similar
manner in the wave packet (2) in the color direction $T^1$, 
i.e. the amplitude $A_1^1$ is decreased. 
The same happens in wave packet (2) but for exchanged indices.
In all six terms appears the longitudinal
amplitude $A_3^{(3)}$ which causes the change of
the color polarization of the receding wave packets. 

The most important role in this context  
play the $g^2$-terms since the longitudinal gauge
field enters quadratically.  
Here, the order $g$ terms on the r.h.s. of Eq (\ref{Eq.47}) act
as a trigger for the $g^2$-term. 
As for example in the case of wave packet (1),
the amplitue $A_2^{(1)}$ is zero initially but grows through 
the corresponding source terms on the r.h.s. of Eq (\ref{Eq.47}).
As long as $A_3^{(3)}$ and $A_2^{(1)}$ are small,
the contribution of the $g^2$-term is suppressed
by the high powers. If these amplitudes exceed a 
certain strength, the $g^2$-term takes over to determine 
the dynamics of $A_2^{(1)}$ resulting in a fast growth. 
In the case of wave packet (1), these growing amplitudes
$A_2^{(1)}$ and $A_3^{(3)}$ enter essentially into
the third term on the r.h.s. of Eq. (\ref{Eq.45})
inducing the growth of $A_3^{(3)}$. 

In order to understand the time delay of the burst-like behavior, 
we study the leading orders in the time dependence of the gauge fields 
shortly after the collision 
\begin{eqnarray}
\label{Eq.48}
A_2^{(1)} &\simeq & a_2^{(1)}t\,+\,b_2^{(1)}t^2\,+\, ...
\\
\label{Eq.49}  
A_3^{(3)} &\simeq & a_3^{(3)}t\,+\,b_3^{(3)}t^2\,+\, ...\, .
\end{eqnarray}
Both amplitueds are zero before the overlapp of the wave packets.
We insert these amplitudes into the Yang-Mills equations
and consider in leading orders 
the effects of the $g^2$-term on the $A_3^{(3)}$.
\begin{eqnarray}
\label{Eq.50}
\partial_t\partial^t A_3^{(3)} 
& \simeq & 
-{{g^2}} A_2^{(1)}A_3^{(3)}A^{(1)}_2\,+\, ... 
\nonumber \\
& = & - {{g^2}} a_2^{(1)}a_3^{(3)}a^{(1)}_2 t^3\,+\, ... 
\end{eqnarray}
We integrate both sides of Eq. (\ref{Eq.50}) in time and
find a leading fourth order time dependence of the longitudinal
color electric field $E_3^{(3)} = -\partial_t A_3^{(3)} $
according to
\begin{eqnarray}
\label{Eq.51}
E_3^{(3)}(t) 
& \simeq & 
{{g^2}\over 4} a_2^{(1)}a_3^{(3)}a^{(1)}_2 t^4\,+\, ... 
\end{eqnarray}
Consequently, the longitudinal field energy increases like 
$ W_L^{(E)}(t) \sim  t^8 $ 
in the leading order of the time dependence.
The high power explains why the longitudinal polarization of the
receding wave packets is small at short times after
the collision but increases rapidly at large times.
This time-dependent behavior of $W_L^{(E)}(t)$ together
with the resulting transverse expansion of the
energy densitiy distribution characterises 
the ''glue burst''. 

\noindent
Our numerical results above show
that the time-delay of the burst scales essentially
like $1/g^2$. This scaling has the following 
analytic explanation. Each factor $a_c^{(\mu)}$ in Eq.
(\ref{Eq.51}) is proportional to $g^2$ and thus
can be separated as 
\begin{eqnarray}
\label{Eq.52}
a_c^{(\mu)} =  \tilde a_c^{(\mu)} g^2  
\end{eqnarray}
where $\tilde a_c^{(\mu)}$ is independent on $g$.
This $g^2$-dependence results from the contribution of the $g^2$-term  
during the overlap time as discussed above. The equation
(\ref{Eq.51}) rereads now
\begin{eqnarray}
\label{Eq.53}
E_3^{(3)}(t) 
& \simeq & 
{1\over 4}\tilde a_2^{(1)}\tilde a_3^{(3)}\tilde a^{(1)}_2 
(g^2 t)^4\,+\, ... 
\end{eqnarray}
which explains the numerical observation, i.e.
$ W_L^{(E)}(t) \sim  (g^2 t)^8 $. 
As Fig. 4 shows, the burst leads to a peak-like shape 
in $W_L^{(E)}(t)$ at the burst time. The steep rise
of the longitudinal energy is immediately followed by a strong
decrease. This behavior has to be explained by the next order
in the expansions (\ref{Eq.48}) and  (\ref{Eq.49}).
The oscillating nature of the solutions suggests that 
the coefficients $b_{\mu}^c$ have opposite sign as compared
to the coefficients $a_{\mu}^c$. This results in a next to leading
term on the r.h.s. of Eq. (\ref{Eq.53}) which has the form
\begin{eqnarray}
\label{Eq.54}
E_3^{(3)}(t) 
& \simeq & 
...\,+\,{1\over {5g^2}}(\tilde b_2^{(1)}\tilde a_3^{(3)}\tilde a^{(1)}_2 
\,+\, ... ) (g^2 t)^5\,+\, ...\,. 
\end{eqnarray} 
Due to the higher power in $t$, this term takes over shortly after
the first term has increased the amplitude of $E_3^{(3)}(t)$. The 
opposite sign however turns the total amplitude back. The following
drop of the amplitude is stopped by an opposite effekt from
the next higher order in Eq. (\ref{Eq.54}) and so on.

\noindent
In Fig. 4, we observe that $ W_L^{(E)}(t) $ stays finite and
seems to oscillate irregularly around a finite average value 
after the burst which we denote preliminary as 
$ \overline W_L^{(E)}(\infty) $.
This average value is independent on $g$ which is explained by
Eq. (\ref{Eq.53}). Eq. (\ref{Eq.53}) shows that in lowest order 
in time, $g^2$ can be completely scaled out with $t$.
As discussed above, the lowest order determines the rise of  
the first hump in $ W_L^{(E)}(t) $ at the beginning of the burst.
The rise of the first hump determines $\overline W_L^{(E)}(\infty) $. 

\noindent
The arguments made in the discussion hold for wave packets of 
large finite transverse extension where we can neglect surface effects.
For small transverse extensions, the discussion 
is in principle similar but more involved due to contributions 
from the surface.
%
%
%
%
\section {Summary and Outlook}
%
%
We have studied time-dependent solutions of the
classical Yang-Mills equations which describe the
collision of initially polarized wave packets
in color space and position space. We have simulated
the collisions on a three dimensional gauge lattice
numerically applying the Hamiltonian approach of
Kogut and Susskind to describe the dynamics of
the color fields in SU(2) gauge symmetry.
As a function of time, we have calculated the
transverse and longitudinal energy densities $w_T^{(E)}(t,x_3)$
and $w_L^{(E)}(t,x_3)$ of the color electric fields.
For initially transverse polarized colliding 
plane wave packtes and for colliding  
finite wave packets as well, the longitudinal 
energy densities show a strongly time dependent
increase in the overlap region around the center of
collision but vanish when the wave packtes recede.
A similar time-dependent behavior was found for
the transverse total energy current. A certain time
$\tau$ after the collision both the longitudinal
color electric field energy and the transverse
energy current increase rapidly while the
distribution $w_T^{(E)}(t,x_3)$ starts simultaneously
to decay. Visualizations in three dimensions
show that the wave packets suddenly decay fast 
in a decoherent manner when the time $\tau$ is reached. 
Both, the maximum transverse energy current at total
overlap and $\tau$ scale like $1/g^2$ but for
different reasons. This and
also the burst can be explained by analyzing the 
Yang-Mills equations.

The question arises, whether this pure classical phenomenon
could play a role in high energy nucleus-nucleus collisions.
In the present calculations each wave packet was
carrying an energy of $\simeq 10\,{\rm GeV}$.
The size of the finite wave packets was $10\, {\rm fm}$
in longitudinal and $8\,{\rm fm}$ in the transverse
directions. The energy of $\simeq 10\,{\rm GeV}$ is
close to the upper limit that can be described on
a lattice with the constant $a_l=0.1\,{\rm fm}$ for
the above size of the wave packet. It is 
well known experimentally that about half of the
energy in a nucleus is carried by glue fields.
A $^{208}Pb$-nucleus at $100\, GeV/u$ carries
thus about $\sim 10\,{\rm TeV}$ in glue fields.
It would be interesting, to study the pure glue field
dynamics classically in colliding finite wave packets
each carrying an energy of $10\,{\rm TeV}$. 
This however requires extremely small lattice constants
and very large numbers of lattice points. 
The size of the wave packets can be adjusted to the size
of colliding $Pb$ nuclei, i.e. it should have an extension of 
about $11\,{\rm fm}$ in the transverse directions.
Such a description of course is still very rough and would require
much improvement in the future.

\noindent
It would also be iteresting to perform a Fourier analysis
of the collisions in three dimensions for each color separately. 
There is hope that Dirac-Fermion fields can be included
in the future.

\noindent
The authors thank S.G. Matinyan for discussions.
This work was supported by the U.S. Department of Energy under
Grant No. DE-FG02-96ER40495.

\newpage
\appendix
%
%
\section {Analysis of the source terms}
\vskip0.5cm
%
We discuss the source terms on the r.h.s. of the
the Yang Mills equations which read
\begin{eqnarray}
\label{A.1}
\partial_{\mu}\partial^{\mu}{\cal A}^{\nu}&=&
\partial_{\mu}\partial^{\nu}{\cal A}^{\mu} \,+\, ig\partial_{\mu}
\bigl[ {\cal A}^{\mu},{\cal A}^{\nu}\bigr]_-  
\nonumber \\
& + &ig\bigl[{\cal A}_{\mu},\partial^{\mu}{\cal A}^{\nu}\bigr]_-
\nonumber \\
& - &ig\bigl[{\cal A}_{\mu},\partial^{\nu}{\cal A}^{\mu}\bigr]_- 
\nonumber \\
& + &g^2\bigl[{\cal A}_{\mu},
        \bigl[{\cal A}^{\mu},{\cal A}^{\nu}\bigr]_-\bigr]_-.  
\end{eqnarray}
Since the time delayed burst-like behavior of the solution
occurs also in the case of
colliding plane waves, we discuss Eq. (\ref{A.1}) first
for this case. 
The numerical results presented in the
section 3 have shown that longitudinal energy densities
correspond to transverse energy currents. The
longitudinal energy densities therefore
exhibit the basic feature of the glue burst solution. The
corresponding color electric field components are determined by the
negative time derivative of the longitudinal components 
of the gauge fields. 
We therefore begin with the discussion of the time evolution
of these field components and we set $\nu=3$ in 
Eq. (\ref{A.1}). Before the collision 
both wave packets are polarized in the
$x_1$-direction in Euclidean space. Accordingly, the
longitudinal components are ${\cal A}^{3} = 0$ before overlap.
In the following we focus on the time region where both wave packets
start to overlap.

The first term on the r.h.s. of Eq. (\ref{A.1}) contains a 
sum of four terms 
\begin{equation}
\label{A.2}
\partial_{\mu}\partial^3{\cal A}^{\mu} =
\partial_{0}\partial^3{\cal A}^{0}     +
\partial_{1}\partial^3{\cal A}^{1}     +
\partial_{2}\partial^3{\cal A}^{2}     +
\partial_{3}\partial^3{\cal A}^{3}.     
\end{equation}
The first term on the r.h.s. of Eq. (\ref{A.2}) is zero because 
we use the temporal gauge in which ${\cal A}^{0}=0$.
The second term and the third term are zero because 
$\partial_{1}{\cal A}^{1}=0$ and $\partial_{2}{\cal A}^{2}=0$ 
for plane waves that propagate into the $x_3$-direction.
For $\nu = 3$, explicitely considering the contribution
from Eq. (\ref{A.2}) on the r.h.s., Eq. (\ref{A.1}) now reads 
\begin{eqnarray}
\label{A.3}
\partial_{0}\partial^{0}{\cal A}^{3}+\partial_{1}\partial^{1}{\cal A}^{3}+
\partial_{2}\partial^{2}{\cal A}^{3}+\partial_{3}\partial^{3}{\cal A}^{3}
&=&
\partial_{3}\partial^{3}{\cal A}^{3}\,+\, ...\quad 
\end{eqnarray}
This shows that the remaining term of the expression
(\ref{A.2}) is canceled by the last term on the l.h.s. of Eq. (\ref{A.3}).
For plane waves, the second and third term on the l.h.s. are
zero. The contributions from the other source terms in 
Eq. (\ref{A.1}) are indicated by the dots.

The second term on the r.h.s of Eq. (\ref{A.1}) is zero
initially because of two reasons. First, ${\cal A}^{3}=0$ at the
very start of the overlapping. The second reason
is explained in the following analysis.
The term in
$\partial_{\mu}\bigl[ {\cal A}^{\mu},{\cal A}^{3} \bigr]_- $
corresponding to $\mu =0$ vanishes due to temporal
gauge, the $\mu = 3$ term vanishes because of the
commutator, and the $\mu = 1,2$ terms vanish 
for plane wave packets.
Consequently, we find
\begin{eqnarray}
\label{A.4} 
\partial_{\mu}\bigl[ {\cal A}^{\mu},{\cal A}^{3} \bigr]_- =0.
\end{eqnarray}

By similar arguments and because of
$\bigl[{\cal A}_{3},\partial^{3}{\cal A}^{3}\bigr]_- = 0$,
we obtain
\begin{eqnarray}
\label{A.5}
ig\bigl[{\cal A}_{\mu},\partial^{\mu}{\cal A}^{3}\bigr]_- =
 \,ig\bigl[{\cal A}_{1},\partial^{1}{\cal A}^{3}\bigr]_- 
\,+\,ig\bigl[{\cal A}_{2},\partial^{2}{\cal A}^{3}\bigr]_- 
\,+\,ig\bigl[{\cal A}_{3},\partial^{3}{\cal A}^{3}\bigr]_- = 0
\end{eqnarray}
for the third term on the r.h.s. of Eq. (\ref{A.1}). 
 
The fourth term on the r.h.s. of Eq. (\ref{A.1}) plays an
important role in the overlap region. 
According to our calculation for plane waves, we assume that the
colliding wave packets are initially polarized in color space
in the directions $T_1$ and $T_2$ respectively. 
When the wave packets
start to overlap, a superposition of two colors occurs and 
we obtain
\begin{eqnarray}
\label{A.6}
-ig\bigl[{\cal A}_{\mu},\partial^3{\cal A}^{\mu}\bigr]_- & = &
-ig\bigl[\bigl( A_{\mu}^1 T_1\,+\, A_{\mu}^2 T_2\,+\, ... \bigr),\partial^3
\bigl( A^{\mu}_1 T^1\,+\, A^{\mu}_2 T^2\,+\, ... \bigr)\bigr]_-  \nonumber \\
& \simeq &
-ig\Bigl( A_{\mu}^1\partial^3 A_2^{\mu} \bigl[T_1, T^2\bigr]_-\,+\,
          A_{\mu}^2\partial^3 A_1^{\mu} \bigl[T_2, T^1\bigr]_-\Bigr) 
\nonumber \\
&=&
{g}\bigl( A_{\mu}^1\partial^3 A_2^{\mu} -
                 A_{\mu}^2\partial^3 A_1^{\mu}\bigr) T_3
\end{eqnarray}
The color indices 1 and 2 refer here at the same time to
the contributions of wave packet 1 and 2. This assumption
is allowed without restriction of the general case in which
the wave packets are polarized in any direction in the
color sub-space ${\rm span}(T_1,T_2)$. 
On the r.h.s. of Eq. \ref{A.6}, we have neglected the terms 
$-ig\bigl[A_{\mu}^3 T_3,\partial^3
 \bigl( A^{\mu}_1 T^1\,+\, A^{\mu}_2 T^2 \bigr)\bigr]_- $ 
and
$ -ig\bigl[\bigl( A_{\mu}^1 T_1\,+\, A_{\mu}^2 T_2 \bigr),\partial^3
  A^{\mu}_3 T^3 \bigr]_- $ 
since $A^{\mu}_3 =0$ at the begin of the overlapping. 
Here, for briefness, we assume that 
$\big\vert A_{\mu}^3  A^{\mu}_3\big\vert \ll
  \big\vert A_{\mu}^1  A^{\mu}_1\big\vert $
and  
$\big\vert A_{\mu}^3  A^{\mu}_3\big\vert \ll
 \big\vert A_{\mu}^2  A^{\mu}_2\big\vert $ during the 
overlap time. 
Consequently, we neglect excitations of modes
in the color sub-space
${\rm span}(T_1,T_2)$ in ${\cal A}^{3}(t,\vec x)$
during the overlap time.  
These components can be included in a 
straight forward manner in the subsequent discussion 
but they do not change the results qualitatively.
What we show does not depend on the above assumption. 
Nevertheless, our numerical results show that the assumption 
is true for many cases.
The remaining term on the r.h.s. in Eq. (\ref{A.6}) acts as a 
source term for the third color component of the 
longitudinal gauge field ${\cal A}^{3}(t,\vec x)$.
Its effect is discussed in section 4.

\noindent
In the following we analyze
the $g^2$-term assuming that the remaining longitudinal
fields are essentially polarized in the $T_3$-direction
in color space. For briefness, we omit the contributions 
from the first and second color direction. This is
sufficient since the $T_3$-component becomes always
finite in the overlap region as shown above.
The $g^2$-term breaks up into
\begin{eqnarray}
\label{A.7}
g^2\bigl[{\cal A}_{\mu},
   \bigl[{\cal A}^{\mu},{\cal A}^{3}\bigr]_-\bigr]_-  
=g^2\bigl[{\cal A}_{1},
     \bigl[{\cal A}^{1},{\cal A}^{3}\bigr]_-\bigr]_-  
 \,+\, g^2\bigl[{\cal A}_{2},
       \bigl[{\cal A}^{2},{\cal A}^{3}\bigr]_-\bigr]_-  
\end{eqnarray}
In the following we discuss the contribution of the
first term on the r.h.s. of Eq. (\ref{Eq.7}).
The second term does not contribute if the wave packets
are polarized into the $x_1$-direction or it
contributes in an analogous manner.
In the following, we put indices in Minkowski-space
into parentheses in order to distinguish from
color indices. 
With the above assumtion, we obtain
\begin{eqnarray}
\label{A.8}
\bigl[{\cal A}^{1},{\cal A}^{3}\bigr]_-
& \simeq &
\bigl[A^{(1)}_1 T^1+A^{(1)}_2T^2+A^{(1)}_3T^3,A^{(3)}_3T^3\bigr]_-
\nonumber \\
& = &
\,-\,{i}A_1^{(1)}A_3^{(3)}T^2\,+\,{i}A_2^{(1)}A_3^{(3)}T^1
\end{eqnarray}
for the inner commutator.
For Eq. (\ref{A.7}), we thus obtain 
\begin{eqnarray}
\label{A.9}
g^2\bigl[{\cal A}_{(1)},
    \bigl[{\cal A}^{(1)},{\cal A}^{(3)}\bigr]_-\bigr]_-  
& = &
\,-\,g^2\bigl[ {\cal A}_{(1)}, {i}A_1^{(1)} A_{3}^{(3)}T^2
                         -{i}A_2^{(1)} A_{3}^{(3)}T^1
   \bigr]_-
\nonumber \\
& = &
\,{{g^2}} A_{(1)}^1A_1^{(1)}A_3^{(3)}\, T^3 
\,-\,{{g^2}} A_{(1)}^3A_1^{(1)}A_3^{(3)}\, T^1 
\nonumber \\
&  &
\,+\,{{g^2}} A_{(1)}^2A_2^{(1)}A_3^{(3)}\, T^3 
\,-\,{{g^2}} A_{(1)}^3A_2^{(1)}A_3^{(3)}\, T^2 
\end{eqnarray}

\noindent
As soon as the wave packets obtain a small contribution
in the gauge field component $A_3^{(3)}$, the component
$A^{(1)}$ starts to change the polarization in color space.
In wave packet (1) the components $A_2^{(1)}$ and $A_3^{(1)}$ 
grow whereas $A_1^{(1)}$ decreases accordingly.    
In wave packet (2) the components $A_1^{(1)}$ and $A_3^{(1)}$ 
grow whereas $A_2^{(1)}$ decreases accordingly.    

\noindent
This behavior is explained by the Yang-Mills equation for
the first component
\begin{eqnarray}
\label{A.10}
\partial_{\mu}\partial^{\mu}{\cal A}^{1}&=&
\partial_{\mu}\partial^{1}{\cal A}^{\mu} \,+\, ig\partial_{\mu}
\bigl[ {\cal A}^{\mu},{\cal A}^{1}\bigr]_-  
\nonumber \\
& + &ig\bigl[{\cal A}_{\mu},\partial^{\mu}{\cal A}^{1}\bigr]_-
\nonumber \\
& - &ig\bigl[{\cal A}_{\mu},\partial^{1}{\cal A}^{\mu}\bigr]_- 
\nonumber \\
& + &g^2\bigl[{\cal A}_{\mu},
        \bigl[{\cal A}^{\mu},{\cal A}^{1}\bigr]_-\bigr]_-.  
\end{eqnarray}
The first term on the r.h.s. of Eq. (\ref{A.10}) does not
contribute for plane waves since $\partial^1{\cal A}^{\mu}=0$.
The second term yields the contributions
\begin{eqnarray}
\label{A.11}
ig\partial_{\mu}\bigl[ {\cal A}^{\mu},{\cal A}^{1}\bigr]_-
& = &
ig\bigl[\partial_3 A_3^{(3)} T^3+...,A_1^{(1)}T^1+A_2^{(1)}T^2+...\bigr]_-
\nonumber\\
&   &
ig\bigl[A_3^{(3)} T^3+...,\partial_3 A_1^{(1)}T^1+
                          \partial_3 A_2^{(1)}T^2+...\bigr]_-
\nonumber\\
& = &
\,-\,g A_1^{(1)}\partial_{(3)} A_3^{(3)} T^2
\,+\, g A_1^{(1)}\partial_{(3)} A_3^{(3)} T^1
\nonumber\\
&  &
\,-\, g A_3^{(3)}\partial_{(3)} A_1^{(1)} T^2
\,+\, g A_3^{(3)}\partial_{(3)} A_2^{(1)} T^1 \,+\, ....
\end{eqnarray}
of which the first and third lead to excitations
of modes in the color direction $T^2$ in the wave packet
(1). The second and fourth term leads to excitations
of modes in the color direction $T^1$.

We now discuss the $g^2$-term. Here, 
$\bigl[ {\cal A}^1,{\cal A}^1\bigr]_- = 0$ in the
inner commutator and ${\cal A}^2=0$ for polarized
wave packets.
For the essential contributions of the inner commutator we find
\begin{eqnarray}
\label{A.12}
\bigl[ {\cal A}^{3},{\cal A}^{1}\bigr]_-
& = &
\bigl[\partial_3 A_3^{(3)} T^3+...,A_1^{(1)}T^1+A_2^{(1)}T^2+...\bigr]_-
\nonumber\\
& = &
i A_3^{(3)} A_1^{(1)} T^2 \,-\, i A_3^{(3)} A_2^{(1)} T^1 \,+\, ....
\end{eqnarray}
Inserting the r.h.s. of Eq. (\ref{A.12}) into the $g^2$-term,
we obtain
\begin{eqnarray}
\label{A.13} 
g^2\bigl[{\cal A}_{\mu},
   \bigl[{\cal A}^{\mu},{\cal A}^{1}\bigr]_-\bigr]_-
&=&
g^2\bigl[{\cal A}_{3},
    \bigl[{\cal A}^{3},{\cal A}^{1}\bigr]_-\bigr]_-  
\nonumber \\
& = &
g^2 A_3^{(3)}A_3^{(3)}A_1^{(1)} T^1
\,+\, g^2 A_3^{(3)}A_3^{(3)}A_2^{(1)} T^2 \,+\, ...\,.  
\end{eqnarray}
%
%
%
\section {Time delay at large momenta}
\vskip0.5cm
%
%
In this appendix, we briefly discuss the $\overline k_3$-scaling in
the overlap region and in the burst region.
As discussed in section 4, for not too large $g$, the fourth 
term in the equation
\begin{eqnarray}
\label{B.1}
\partial_{\mu}\partial^{\mu}{\cal A}^{\nu}&=&
\partial_{\mu}\partial^{\nu}{\cal A}^{\mu} \,+\, ig\partial_{\mu}
\bigl[ {\cal A}^{\mu},{\cal A}^{\nu}\bigr]_-  
\nonumber \\
& + &ig\bigl[{\cal A}_{\mu},\partial^{\mu}{\cal A}^{\nu}\bigr]_-
\nonumber \\
& - &ig\bigl[{\cal A}_{\mu},\partial^{\nu}{\cal A}^{\mu}\bigr]_- 
\nonumber \\
& + &g^2\bigl[{\cal A}_{\mu},
        \bigl[{\cal A}^{\mu},{\cal A}^{\nu}\bigr]_-\bigr]_-  
\end{eqnarray}
acts as the dominant source term for the longitudinal
field components in the overlap region
of the colliding wave packets.
We treat Eq. (\ref{B.1}) first for $\nu = 3$.
Assuming that during the overlap time the wave packets 
deviate not much from their initial form
\begin{eqnarray}
\label{B.2}
\phi(t,x_3)&=&\phi_0
\exp{\bigl(-{1\over 2}\Delta k_3^2(t+x_3)^2\bigr)}
\cos{\bigl(\overline k_3(t+x_3)\bigr)},
\end{eqnarray}
we can make the approximation 
$ \partial^{3}{\cal A}^{\mu} \simeq \overline k_3 {\cal A}^{\mu} $
at large $\overline k_3$.
Consequently, the amplitude ${\cal A}^{3} $ is proportional
to $\overline k_3$ during the overlap time. This leads to
a $1/{\overline k_3^2}$-scaling of the hight of the hump 
in $w_L^{(E)}(t,z)$ in the overlap region. 
Further, ${\cal A}^{3} \sim \overline k_3 $ enters into
the $g^2$-term on the r.h.s. of Eq. (\ref{B.1}).
This leads to a linear $\overline k_3$-dependence of
the coefficients $\tilde a_c^{\mu}$ which appear in Eq. (\ref{Eq.53})
in the section 4. Separating $\overline k_3$ from the coefficients
leads to
\begin{eqnarray}
\label{B.3}
E_3^{(3)}(t) 
& \simeq & 
{1\over 4}{\overline a}_2^{(1)}{\overline a}_3^{(3)}{\overline a}^{(1)}_2 
(g^2 {\overline k_3}^{3/4} t)^4\,+\, ... 
\end{eqnarray} 
with new coefficients 
${\overline a}_c^{\mu} = \tilde a_c^{\mu}/{\overline k_3}$.
We conclude that the time delay of the burst scales like
${\overline k_3}^{-3/4}$. 
%
%


\begin{thebibliography}{999}
\bibitem{Bass.98} S.A. Bass et al., Prog. Part. Nucl. Phys. {\bf 41}
                  (1998) 225-370. 
\bibitem{Faessler.98} S. Ben-Hao et al., nucl-th/9809020.
\bibitem{Venu.97} A. Krasnitz and R. Venugopalan, hep-ph/9706329 
                  (March 1998).
\bibitem{Wang.96} X.-N. Wang, Phys. Rept. {\bf 280}, 237 (1996).
\bibitem{Geiger.95} K. Geiger, Phys. Rept. {\bf 258}, 237 (1995).
\bibitem{Geiger.92} K. Geiger and B. M\"uller, {\em Nucl. Phys.}
                    {\bf B369}, 600 (1992); K. Geiger, hep-ph/9701226.
\bibitem{McLerran.94} L. McLerran and R. Venugopalan, {\em Phys. Rev.}
                      {\bf D49}, 2233 and 3352 (1994)
\bibitem{BMP.98} S.A. Bass, B. M\"uller, and W. P\"oschl, Duke University
                  preprint DUKE-TH-98-168 (nucl-th/9808011).
\bibitem{PM.98} W. P\"oschl, B. M\"uller, Duke University
                  preprint DUKE-TH-98-169 (nucl-th/9808031).
\bibitem{Hu.95} C.R. Hu, et al.,
                {\em Phys. Rev. D} {\bf 52}, 2402 (1995).
\bibitem{Ring.91} A. Ringwald, {\em Nucl. Phys.} {\bf B330}, 1 (1989).
\bibitem{Esp.91} O. Espinosa, {\em Nucl. Phys.} {\bf B343}, 310 (1990).
\bibitem{Gong.94} C. Gong, S.G. Matinyan, B. M\"uller and A. Trayanov
                  {\em Phys. Rev.} {\bf D49}, 607 (1994).

\end{thebibliography}
\end{document}